\documentclass[12pt,a4paper]{article}
\usepackage{amssymb,enumitem,booktabs,subfig,xcolor,microtype,setspace,bm,longtable}
\usepackage[top=1.4in, bottom=1.4in, left=1.35in, right=1.35in]{geometry}
\usepackage{natbib}
\usepackage{url} 
\usepackage[pdftex,colorlinks=true,hypertexnames=false]{hyperref}
\definecolor{darkblue}{rgb}{0,0,.6}
\hypersetup{citecolor=darkblue,linkcolor=darkblue,urlcolor=darkblue}
\usepackage{amsmath}
\usepackage{amsfonts}
\usepackage{epsfig}
\usepackage{graphics}
\usepackage[normalem]{ulem}
\usepackage{palatino,eulervm}
\usepackage{graphicx}
\usepackage{lscape}
\usepackage{rotating}
\usepackage[linewidth=1pt]{mdframed}
\allowdisplaybreaks[4]
\DeclareMathOperator*{\argmin}{arg\,min}

\providecommand{\U}[1]{\protect\rule{.1in}{.1in}}

\graphicspath{{plots/}}

\usepackage{amsthm,thmtools}
\usepackage{mathrsfs}
\declaretheorem{theorem}
\declaretheorem{lemma}
\makeatletter
\def\th@newremark{\th@remark\thm@headfont{\bfseries}}
\makeatletter
\theoremstyle{newremark}
\newtheorem{remark}{Remark}
\newtheorem{prop}{Proposition}

\newtheorem{assumption}{Assumption}
\declaretheoremstyle[
  spaceabove=6pt, spacebelow=6pt,
  headfont=\bfseries,
  notefont=\mdseries, notebraces={(}{)},
bodyfont=\normalfont,
  postheadspace=0.5em,
]{mystyle}

\begin{document}

\title{Covariance Function Estimation for High-Dimensional Functional Time Series with Dual Factor Structures}
\author{{\normalsize Chenlei Leng\thanks{%
Department of Statistics, University of Warwick, UK},\ \ \ Degui Li\thanks{%
Department of Mathematics, University of York, UK. },\ \ \ Han Lin Shang\thanks{Department of Actuarial Studies and Business Analytics, Macquarie University, Australia.},\ \ \ Yingcun Xia\thanks{Department of Statistics and Data Science, National University of Singapore, Singapore.}}\\
{\normalsize\em University of Warwick,\ \ University of York,\ \ Macquarie University,\ \ National University of Singapore}}

\date{\normalsize This version: \today}

\maketitle

\centerline{\bf Abstract}

\medskip

We propose a flexible dual functional factor model for modelling high-dimensional functional time series.  In this model, a  high-dimensional fully functional factor parametrisation is imposed on the observed functional processes, whereas a low-dimensional version (via series approximation) is assumed for the latent functional factors. We extend the classic principal component analysis technique for the estimation of a low-rank structure to the estimation of a large covariance matrix of random functions that satisfies a notion of (approximate) functional ``low-rank plus sparse" structure; and generalise the matrix shrinkage method to functional shrinkage in order to estimate the sparse structure of functional idiosyncratic components. Under appropriate regularity conditions, we derive the large sample theory of the developed estimators, including the consistency of the estimated factors and functional factor loadings and the convergence rates of the estimated matrices of covariance functions measured by various (functional) matrix norms. Consistent selection of the number of factors and a data-driven rule to choose the shrinkage parameter are discussed. Simulation and empirical studies are provided to demonstrate the finite-sample performance of the developed model and estimation methodology.

\medskip

\noindent{\em Keywords}: Covariance operator, functional factor model, functional time series, generalised shrinkage, high dimensionality, PCA, sparsity.

\newpage


\section{Introduction}\label{sec1}
\renewcommand{\theequation}{1.\arabic{equation}}
\setcounter{equation}{0}

A fundamental problem of increasing interest in modelling time series of random functions is to estimate their second-order characteristics, such as the covariance, auto-covariance and spectral density operators \citep[e.g.,][]{B00, HK10, HK12}. Understanding these characteristics are not only important for understanding the randomness of the corresponding processes themselves, but also crucial to subsequent down-stream applications such as functional principal component analysis \citep[e.g.,][]{HK10, HK12, PT13, HKH15}. While most of the existing literature focuses on a fixed number of functional time series,  it is increasingly common to collect a large number of them in a diverse range of fields. For example, in climatology, temperature curves are routinely recorded in hundreds of weather stations, while in finance, stock return curves are typically available for thousands of stocks. 

The main aim of this paper is to estimate the covariance structure of functional time series, when the dimensionality of the data (the number of functional time series) is comparable to or greater than the sample size (the length of these time series). To overcome the curse of dimensionality arising from this setup, a common practice is to make an approximate sparsity assumption on this structure. Consequently, regularisation methods such as thresholding or adaptive thresholding can be applied  \citep[e.g.,][]{FGQ23}. However, the sparsity restriction only works when these functional time series are at most weakly correlated, ruling out many interesting cases where they can be highly correlated. In practice, it is widely recognised that multivariate functional time series are often influenced by common functions over the temporal dimension, leading to strong cross-sectional dependence. For example, the rainfall curves collected in many stations may be affected by the common weather pattern in the region \citep{DHP19}, and the intraday return curves are often driven by latent market and industry factors.

To accommodate cross-sectional dependence, an idea is to use factor models in the context of functional time series. Indeed, the approximate factor model has been extensively studied in the literature for panel time series \citep{CR83, BN02, SW02, FLM13}. In recent years, there have been some attempts to extend this model to the functional time series setting, broadly classified into two directions. The first category can be loosely referred to as functional factor models for fixed-dimensional problems. Specifically, \cite{SH12} and \cite{KMZ15} consider a factor model with real-valued factor loadings and functional factors, whereas \cite{KMRT18} and \cite{MGG22} propose a factor model with real-valued factors and functional loadings. These functional factor models are essentially {\em low-dimensional}, because the developed methodology is not applicable to large-scale functional time series when the dimension is comparable to the sample size. The second direction is the {\em high-dimensional} functional factor model. \cite{GSY19} first apply the classic functional principal component analysis to each functional time series before modelling the component scores via a factor model. However, this method may lead to information loss in the dimension reduction stage, which subsequently may result in inaccurate factor number estimation. To address these, two types of functional factor models with different construction of common components have been proposed in the recent literature. \cite{TNH23a, TNH23b} introduce a high-dimensional functional factor model with functional loadings and real-valued factors, whereas \cite{GQW21} propose a different version with real-valued loadings and functional factors.

In this paper, we propose a novel dual functional factor model with functional loadings and functional latent factors, in which the latent factors are modelled through series approximation. Towards this, we relax some restrictive assumptions imposed on existing functional factor models studied in the aforementioned literature. The proposed model includes those in \cite{TNH23a, TNH23b} as special cases. We allow the covariance functions for the idiosyncratic components to be approximately sparse in a functional sense. Thus, the covariance function (or operator) of the high-dimensional functional process under our setup admits an approximate ``low-rank plus sparse" functional covariance structure \citep[e.g.,][]{FLM13}, extending the functional sparsity condition assumed in \cite{FGQ23} to allow latent factors.

The main estimation methodology can be seen as a functional analog of the POET method introduced by \cite{FLM13}. For the factor loadings and common factors, we propose the use of a functional version of principal component analysis (PCA). For the covariance function of the idiosyncratic components, we apply a functional generalised shrinkage to a preliminary estimate of it. Under regularity conditions, we derive the mean squared convergence of the estimated factors, uniform convergence of the estimated functional factor loadings, and convergence of the estimated functional covariance matrices (measured by different matrix norms). These convergence rates depend on the dimension, the length of functional time series, the number of factors, and the error of sieve functional approximation. The Monte-Carlo simulation study shows  that the developed methodology has reliable finite-sample performance. An empirical application to the cumulative intraday returns (CIDR) curves of the S\&P 500 index data confirms the usefulness of the model and the accuracy of the suggested estimation approach.

After completing a preliminary draft of the paper, we found a recent working paper by \cite{LQW23} who consider a similar problem. It is worth comparing the two papers before concluding the introductory section. First, \cite{LQW23} estimate the large functional covariance matrix based on the high-dimensional functional factor models of \cite{TNH23a, TNH23b} and \cite{GQW21}, whereas our estimation is built on the dual functional factor model structure. More discussion and comparison between the models are provided in Section \ref{sec2} below. Second, we allow the number of factors (via the functional sieve approximation) to be divergent and the functional observations from different subjects to be defined on different domains, relaxing some restrictions implicitly imposed in \cite{LQW23}. Third, we derive general convergence theory under the flexible model framework and the obtained convergence rates are comparable to those in \cite{LQW23} where the factor number is fixed and the sieve approximation error is zero. Finally, we use a different criterion to consistently estimate the factor number and choose the tuning parameter in functional shrinkage. In particular, we propose a modified cross-validation to select the shrinkage parameter, taking into account the temporal dependence of the functional data over a long time span.


The rest of the paper is organised as follows. Section \ref{sec2} introduces the dual functional factor model framework. Section \ref{sec3} describes the main estimation methodology and Section \ref{sec4} presents the large sample theory for the developed estimators. Section \ref{sec5} discusses practical issues for implementation and reports the numerical studies. Section \ref{sec6} concludes the paper. Proofs of the main theorems are given in Appendix A and proofs of some technical lemmas are available in Appendix B. Appendix C briefly reviews the $\tau$-mixing dependence and the concentration inequality. Throughout the paper, for a separable Hilbert space ${\mathscr H}$ defined as a set of real measurable functions $x(\cdot)$ on a compact set ${\mathbb C}$ such that $\int_{\mathbb C} x^2(u)du<\infty$, the inner product of $x_1,x_2\in{\mathscr H}$ is $\langle x_1,x_2\rangle=\int_{\mathbb C}x_1(u)x_2(u)du$, and the norm of $x\in{\mathscr H}$ is $\Vert x\Vert_2=\langle x,x\rangle^{1/2}$. Denote by ${\mathscr H}_1\times {\mathscr H}_2$ the Cartesian product of ${\mathscr H}_1$ and ${\mathscr H}_2$. Let $\Vert \cdot\Vert_O$ and $\Vert \cdot\Vert_{\rm S}$ respectively denote the operator and Hilbert-Schmidt norms for continuous linear operators. Denote the Euclidean norm of a vector by $\vert \cdot\vert_2$, and the operator, Frobenius and maximum norms of a matrix by $\Vert \cdot\Vert$, $\Vert\cdot\Vert_F$ and $\Vert\cdot\Vert_{\max}$, respectively. For $A=(a_1,\cdots, a_p)\in{\mathscr H}^p$ (the $p$-fold Cartesian product of ${\mathscr H}$), we define $|||A|||_2^2=\sum_{i=1}^p\int_{\mathbb C}a_i^2(u)du$. Let $a_{n}\sim b_{n}$, $a_{n}\propto b_{n}$ and $a_{n}\gg b_{n}$ denote that $a_{n}/b_{n}\rightarrow1$, $0<\underline{c}\leq a_{n}/b_{n}\leq\overline{c}<\infty$ and $b_{n}/a_{n}\rightarrow0$, respectively. Write ``with probability approaching one" as ``{\em w.p.a.1}" for brevity.


\section{Dual functional factor models}\label{sec2}
\renewcommand{\theequation}{2.\arabic{equation}}
\setcounter{equation}{0}

Let $X_t=(X_{1t},\cdots,X_{Nt})^{^\intercal}$, $t=1,\cdots,T$, with $X_{it}=\left(X_{it}(u):\ u\in{\mathbb C}_i\right)\in{\mathscr H}_i$, where ${\mathscr H}_i$ is a Hilbert space defined as a set of measurable and square-integrable functions on a bounded set ${\mathbb C}_i$. Throughout the paper, we use $i$ to denote the subject index and $t$ to denote the time index. We propose the following functional factor model:
\begin{equation}\label{eq2.1}
X_{it}=\sum_{k=1}^{k_\ast} {\mathbf B}_{ik}F_{tk}+\varepsilon_{it}=:\chi_{it}+\varepsilon_{it},\ \ i=1,\cdots,N,\ \ t=1,\cdots,T,
\end{equation}
where $F_{tk}=\left(F_{tk}(u):\ u\in{\mathbb C}_k^\ast\right)\in{\mathscr H}_k^\ast$, ${\mathbb C}_k^\ast$ is a bounded set which may be different from ${\mathbb C}_i$, ${\mathscr H}_k^\ast$ is defined similarly to ${\mathscr H}_i$ but replacing ${\mathbb C}_i$ by ${\mathbb C}_k^\ast$, ${\mathbf B}_{ik}$ is a continuous linear operator from ${\mathscr H}_k^\ast$ to ${\mathscr H}_i$, and $\varepsilon_{it}=\left(\varepsilon_{it}(u):\ u\in{\mathbb C}_i\right)\in{\mathscr H}_i$. Neither the factor loading operator ${\mathbf B}_{ik}$ nor functional factor $F_{kt}$ is observable. The number $k_\ast$ is unknown but assumed to be finite. Model (\ref{eq2.1}) is a natural extension of the conventional factor model to a panel of functional time series. It follows from (\ref{eq2.1}) that each functional observation $X_{it}$ is decomposed into the functional common and idiosyncratic components: $\chi_{it}$ and $\varepsilon_{it}$. Write $F_t=(F_{t1},\cdots,F_{tk_\ast })^{^\intercal}$, $\varepsilon_t=(\varepsilon_{1t},\cdots,\varepsilon_{Nt})^{^\intercal}$, $\chi_t=(\chi_{1t},\cdots,\chi_{Nt})^{^\intercal}$ and
\[
{\mathbf B}=\left(
\begin{array}{ccc}
{\mathbf B}_{11}&\cdots&{\mathbf B}_{1k_\ast}\\
\vdots&\ddots&\vdots\\
{\mathbf B}_{N1}&\cdots&{\mathbf B}_{Nk_\ast}
\end{array}
\right),
\]
an $N\times k_\ast$ matrix of continuous linear operators from ${\mathscr H}_F:={\mathscr H}_1^\ast\times\cdots\times {\mathscr H}_{k_\ast}^\ast$ to ${\mathscr H}_X:={\mathscr H}_1\times\cdots\times {\mathscr H}_N$. Then model (\ref{eq2.1}) can be written in the following vector/matrix form:
\begin{equation}\label{eq2.2}
X_t={\mathbf B}F_t+\varepsilon_t=\chi_t+\varepsilon_t,\ \ t=1,\cdots,T.
\end{equation}

To facilitate comparison with the functional factor models proposed in the recent literature, we next consider ${\mathbf B}_{ik}$ as a linear integral operator with kernel $B_{ik}=\left(B_{ik}(u,v):\ u\in{\mathbb C}_i, v\in{\mathbb C}_k^\ast\right)$, i,e.,
\[
{\mathbf B}_{ik}(x)(u)=\int_{{\mathbb C}_k^\ast} B_{ik}(u,v)x(v)dv,\ \ x\in{\mathscr H}_k^\ast.
\]
Model (\ref{eq2.1}) can be re-written as
\begin{equation}\label{eq2.3}
X_{it}(u)=\sum_{k=1}^{k_\ast}\int_{{\mathbb C}_k^\ast}B_{ik}(u,v)F_{tk}(v)dv+\varepsilon_{it}(u)=\chi_{it}(u)+\varepsilon_{it}(u),\ \ u\in{\mathbb C}_i.
\end{equation}

\cite{TNH23a} introduce the following model:
\begin{equation}\label{eq2.4}
X_{it}(u)=\sum_{k=1}^{k_\dag}B_{ik}^\dag(u)F_{tk}^\dag+\varepsilon_{it}(u),\ \ u\in{\mathbb C}_i,
\end{equation}
where $B_{ik}^\dag(\cdot)$ is the functional factor loading and $F_t^\dag=\left(F_{t1}^\dag,\cdots,F_{tk_\dag}^\dag\right)^{^\intercal}$ is a vector of real-valued factors. It is easy to see that model (\ref{eq2.4}) is a special case of (\ref{eq2.3}). Letting the matrix ${\mathbf B}$ of operators in (\ref{eq2.2}) be replaced by a matrix of real-valued factor loadings, we obtain the model in \cite{GQW21}:
\begin{equation}\label{eq2.5}
X_{it}(u)=\sum_{k=1}^{k_\ddag}B_{ik}^\ddag F_{tk}^\ddag(u)+\varepsilon_{it}(u),
\end{equation}
where $B_{ik}^\ddag$ is the real-valued factor loading and $F_{tk}^\ddag(\cdot)$ is the functional factor. The factor numbers $k_\ast$ in \eqref{eq2.3} and $k_\ddag$ in \eqref{eq2.4} are assumed to be fixed positive integers. As in \cite{HG18} and \cite{TNH23a, TNH23b}, we allow the functional observations $X_{it}(\cdot)$, $i=1,\cdots,N$, and the latent factors $F_{tk}(\cdot)$, $k=1,\cdots,k_\ast$, to be defined on different domains, i.e., ${\mathbb C}_i$ and ${\mathbb C}_k^\dag$ may vary over $i$ and $k$. For instance, the collected high-dimensional functional time series may be a combination of time series with values in spaces with different dimensions (for different subjects). In contrast, \cite{GQW21} implicitly assume that $X_{it}(\cdot)$ and $F_{tk}^\ddag(\cdot)$ are defined on the same domain.

For the model in (\ref{eq2.1})--(\ref{eq2.3}), we further impose a low-dimensional functional factor condition on the latent factor $F_t$ by assuming the following series approximation:
\begin{equation}\label{eq2.6}
F_{tk}(u)=\Phi_k(u)^{^\intercal}G_t+\eta_{tk}(u),\ \ u\in{\mathbb C}_k^\ast,
\end{equation}
where $\Phi_k(\cdot)$ is a $q$-dimensional vector of basis functions, $G_t$ is a $q$-dimensional vector of stationary random variables, $\eta_{tk}\in{\mathscr H}_k^\ast$ is the sieve approximation error satisfying Assumption \ref{ass:1}(iii), and $q$ is a positive integer which may slowly diverge to infinity. Note that the set of basis functions in (\ref{eq2.6}) is allowed to vary over $k$. Model (\ref{eq2.6}) extends the models studied in \cite{KMRT18} and \cite{MGG22} for univariate functional time series to multivariate ones.

Letting
\[\Lambda_i(u)=\sum_{k=1}^{k_\ast} {\mathbf B}_{ik}\Phi_k(u)=\sum_{k=1}^{k_\ast}\int_{{\mathbb C}_k^\ast}B_{ik}(u,v)\Phi_k(v)dv\]
and
\[
\chi_{it}^\eta(u)=\sum_{k=1}^{k_\ast}{\mathbf B}_{ik}\eta_{tk}(u)=\sum_{k=1}^{k_\ast}\int_{{\mathbb C}_k^\ast}B_{ik}(u,v)\eta_{tk}(v)dv,\]
the dual functional factor model structure combining (\ref{eq2.3}) and (\ref{eq2.6}) leads to
\begin{eqnarray}
X_{it}(u)&=&\Lambda_{i}(u)^{^\intercal}G_{t}+\chi_{it}^\eta(u)+\varepsilon_{it}(u)\nonumber\\
&=&\chi_{it}^\ast(u)+\chi_{it}^\eta(u)+\varepsilon_{it}(u)\nonumber\\
&=&\chi_{it}^\ast(u)+\varepsilon_{it}^\ast(u),\ \ u\in{\mathbb C}_i,\label{eq2.7}
\end{eqnarray}
where $\chi_{it}^\ast(u)=\Lambda_{i}(u)^{^\intercal}G_{t}$ and $\varepsilon_{it}^\ast(u)=\chi_{it}^\eta(u)+\varepsilon_{it}(u)$. By Assumption \ref{ass:1}(ii)--(iii) in Section \ref{sec4} below and the Cauchy-Schwarz inequality, we may show that $\chi_{it}^\eta$, the common component driven by the series approximation error, converges to zero {\em w.p.a.1}. If $\eta_{tk}\equiv0$, model (\ref{eq2.7}) reduces to (\ref{eq2.4}) as in \cite{TNH23a, TNH23b} but with a possibly diverging number of factors. Our main interest is to estimate the large contemporaneous functional covariance structure under the general functional factor model framework (\ref{eq2.7}).


\section{Estimation methodology}\label{sec3}
\renewcommand{\theequation}{3.\arabic{equation}}
\setcounter{equation}{0}

Without loss of generality, assume that $X_{it}, \varepsilon_{it}, F_{kt}$ and $G_t$ all have zero mean. Define ${\mathbf C}_X={\sf E}\left(X_tX_t^{^\intercal}\right)=\left({\mathbf C}_{X,ij}\right)_{N\times N}$, where ${\mathbf C}_{X,ij}={\sf E}\left(X_{it}X_{jt}\right)$ is the covariance operator between $X_{it}$ and $X_{jt}$ with the corresponding kernel $C_{X,ij}(u,v)={\sf E}\left[X_{it}(u)X_{jt}(v)\right]$. Assuming $\chi_t$ and $\varepsilon_t$ are uncorrelated, we may decompose ${\mathbf C}_X$ as
\begin{equation}\label{eq3.1}
{\mathbf C}_X={\mathbf C}_\chi+{\mathbf C}_\varepsilon,\ \ {\mathbf C}_\chi=\left({\mathbf C}_{\chi,ij}\right)_{N\times N},\ \ {\mathbf C}_\varepsilon=\left({\mathbf C}_{\varepsilon,ij}\right)_{N\times N},
\end{equation}
where ${\mathbf C}_{\chi,ij}$ is the covariance operator between the functional common components $\chi_{it}$ and $\chi_{jt}$, and ${\mathbf C}_{\varepsilon,ij}$ is the covariance operator between the functional idiosyncratic components $\varepsilon_{it}$ and $\varepsilon_{jt}$\footnote{Throughout the paper, we use ${\mathbf C}_{X,ij}, {\mathbf C}_{\chi,ij}$ and ${\mathbf C}_{\varepsilon,ij}$ to denote the operators, and $C_{X,ij}(u,v), C_{\chi,ij}(u,v)$ and $C_{\varepsilon,ij}(u,v)$ to denote their respective kernels.}. Let $C_X=\left(C_{X,ij}\right)_{N\times N}, C_\chi=\left(C_{\chi,ij}\right)_{N\times N}$ and $C_\varepsilon=\left(C_{\varepsilon,ij}\right)_{N\times N}$ be matrices of covariance functions (or kernels) corresponding to ${\mathbf C}_X$, ${\mathbf C}_\chi$ and ${\mathbf C}_\varepsilon$, respectively.

It follows from (\ref{eq2.7}) that $C_{\chi,ij}(u,v)={\sf E}[\chi_{it}(u)\chi_{jt}(v)]$ can be approximated by
\begin{equation}\label{eq3.2}
C_{\chi,ij}^\ast(u,v)=\Lambda_i(u)^{^\intercal}\Sigma_G\Lambda_j(v),
\end{equation}
where $\Sigma_G$ is a $q\times q$ covariance matrix of latent factors $G_t$ and $\Lambda_i(u)$ is defined as in (\ref{eq2.7}). Hence $C_\chi$ can be approximated by $C_\chi^\ast=\left(C_{\chi,ij}^\ast\right)_{N\times N}$, a low-rank functional covariance matrix.
On the other hand, since the functional idiosyncratic components are weakly cross-sectionally correlated, it is sensible to impose a functional version of the approximate sparsity restriction on $C_\varepsilon$, i.e., \begin{equation}\label{eq3.3}
C_\varepsilon\in{\mathcal S}(\iota, \varpi_N)=\left\{\Sigma=(\Sigma_{ij})_{N\times N}:\ \Sigma\succeq0,\ \max_{1\leq i\leq N}\sum_{j=1}^N \Vert \Sigma_{ij}\Vert_{\rm S}^{\iota}\le  \varpi_N\right\},\quad0\leq \iota<1,
\end{equation}
where $\varpi_N$ denotes a positive number which depends on $N$, and ``$\Sigma\succeq0$" denotes that $\Sigma$ is a positive semi-definite matrix of covariance functions, i.e.,
$$\sum_{i=1}^N\sum_{j=1}^N \int_{u\in{\mathbb C}_i}\int_{v\in{\mathbb C}_j}x_i(u)\Sigma_{ij}(u,v)x_j(v)dudv\geq0,\ \ \forall\ x=(x_1,\cdots,x_N)^{^\intercal}\in{\mathscr H}_X.$$
Combining the above arguments, we obtain the functional low-rank plus sparse structure for
\begin{equation}\label{eq3.4}
C_X^\ast=\left(C_{X,ij}^\ast\right)_{N\times N}=C_\chi^\ast+C_\varepsilon,
\end{equation}
which may serve as a proxy of $C_X$. Under some mild conditions such as Assumption \ref{ass:1}(iii), we may show $C_X^\ast$ converges to $C_X$ (under the matrix maximum norm), using Lemma \ref{le:B.3}. With (\ref{eq3.4}), we may extend the POET method in \cite{FLM13} to estimate $C_X^\ast$ in the high-dimensional functional data setting.

Due to the general functional structure for the common components in (\ref{eq2.1})--(\ref{eq2.3}), it is practically infeasible to directly estimate the factor loading operators and functional factors in the latent structure. Instead, motivated by (\ref{eq3.2})--(\ref{eq3.4}), we may make use of the low-dimensional functional factor model (\ref{eq2.6}) and estimate $\Lambda_i(\cdot)$ and $G_t$ (subject to rotation) in the latent $\chi_{it}^\ast$ which approximates $\chi_{it}$. The latter can be done by extending the PCA technique \citep[e.g.,][]{BN02, SW02} to a large panel of functional observations. Assume the number of real-value factors in (\ref{eq2.7}) is known for the time being. We will discuss on how to determine this number in Section \ref{sec5}. We estimate the factor loading functions and real-valued factors by minimising the following least squares objective function:
\[
\sum_{t=1}^T \left\Vert X_{t}-\lambda g_t\right\Vert_{N,2}^2=\sum_{t=1}^T\sum_{i=1}^N \left\Vert X_{it}-\lambda_i g_t\right\Vert_2^2,
\]
where $\Vert\cdot\Vert_{N,2}$ denotes the norm of elements in ${\mathscr H}_X$, $\lambda_i=\left(\lambda_{i}(u):\ u\in{\mathbb C}_i\right)$ is a $q$-dimensional vector of functions and $g_t$ is a $q$-dimensional vector of numbers. Minimisation of the above least squares objective function can be achieved via the eigenanalysis of
\begin{equation}\label{eq3.5}
\Delta=\left(\Delta_{ts}\right)_{T\times T}\ \ {\rm with}\ \ \Delta_{ts}=\frac{1}{N}\sum_{i=1}^N \int_{u\in{\mathbb C}_i}X_{it}(u)X_{is}(u)du.
\end{equation}
Consider the following identification condition in the PCA algorithm:
\begin{equation}\label{eq3.6}
\frac{1}{T}\sum_{t=1}^TG_tG_t^{^\intercal}=I_{q}\ \ {\rm and}\ \ \frac{1}{N}\sum_{i=1}^N \int_{u\in{\mathbb C}_i}\Lambda_i(u)\Lambda_i(u)^{^\intercal}du\ \ {\rm is\ diagonal},
\end{equation}
which are similar to those in \cite{BN02} and \cite{FLM13}. With (\ref{eq3.6}), we let $\widetilde G=\big(\widetilde G_1,\cdots,\widetilde G_T\big)^{^\intercal}$ as a matrix consisting of the eigenvectors (multiplied by root-$T$) corresponding to the $q$ largest eigenvalues of $\Delta$ defined in (\ref{eq3.5}). The factor loading functions are estimated as
\[\widetilde\Lambda_i=\left(\widetilde\Lambda_i(u): u\in{\mathbb C}_i\right)=\frac{1}{T}\sum_{t=1}^TX_{it}\widetilde G_t,\ \ i=1,\cdots,N,\]
via the least squares, using the normalisation restriction $\frac{1}{T}\sum_{t=1}^T\widetilde G_t\widetilde G_t^{^\intercal}=I_q$ by (\ref{eq3.6}). Consequently, the low-rank matrix of covariance functions $C_\chi^\ast$ is estimated by
\begin{equation}\label{eq3.7}
\widetilde C_\chi=\left(\widetilde C_{\chi,ij}\right)_{N\times N}\ \ {\rm with}\ \ \widetilde C_{\chi,ij}=\left(\widetilde C_{\chi,ij}(u,v)=\widetilde\Lambda_i(u)^{^\intercal}\widetilde\Lambda_j(v):\ u\in{\mathbb C}_i,\ v\in{\mathbb C}_j\right).
\end{equation}

We next turn to the estimation of $C_\varepsilon$. Letting $
\widetilde\varepsilon_{it}=X_{it}-\widetilde\Lambda_{i}^{^\intercal}\widetilde G_t$ be the approximation of $\varepsilon_{it}$, it is natural to estimate $C_\varepsilon$ by
\[
\widehat{C}_\varepsilon=\left(\widehat{C}_{\varepsilon,ij}\right)_{N\times N}\ \ {\rm with}\ \
\widehat{C}_{\varepsilon,ij}=\left(\widehat C_{\varepsilon,ij}(u,v):\ u\in{\mathbb C}_i,\ v\in{\mathbb C}_j\right)=\frac{1}{T}\sum_{t=1}^T \widetilde\varepsilon_{it}\widetilde\varepsilon_{jt}.
\]
However, this matrix of conventional sample covariance functions often performs poorly when the number $N$ is comparable to or larger than $T$. To address this problem, we adopt the generalised shrinkage and estimate $C_\varepsilon$ by
\begin{equation}\label{eq3.8}
\widetilde{C}_\varepsilon=\left(\widetilde{C}_{\varepsilon,ij}\right)_{N\times N}\ \ {\rm with}\ \
\widetilde{C}_{\varepsilon,ij}=\left(\widetilde C_{\varepsilon,ij}(u,v):\ u\in{\mathbb C}_i,\ v\in{\mathbb C}_j\right)=s_{\rho}\left(\widehat{C}_{\varepsilon,ij}\right),
\end{equation}
where $s_\rho$ is a functional thresholding operator satisfying (i) $\Vert s_\rho(C)\Vert_{\rm S} \leq \Vert C\Vert_{\rm S}$ for any covariance function (or operator) $C$; (ii) $\Vert s_\rho(C)\Vert_{\rm S}=0$ if $\Vert C\Vert_{\rm S} \leq \rho$; and (iii) $\Vert s_\rho(C)-C\Vert_{\rm S}\leq \rho$, where $\rho$ is a user-specified tuning parameter controlling the level of shrinkage. A modified cross-validation method will be given in Section \ref{sec5} to determine $\rho$, taking into account the temporal dependence of high-dimensional functional data. We may further replace the universal thresholding by an adaptive functional thresholding as suggested by \cite{FGQ23} and \cite{LQW23}, in which case the theory to be developed in Section \ref{sec4} remains valid.

Combining $\widetilde{C}_\chi$ and $\widetilde{C}_\varepsilon$, we finally obtain the estimate of $C_X^\ast$ or $C_X$:
\begin{equation}\label{eq3.9}
\widetilde{C}_X=\left(\widetilde{C}_{X,ij}\right)_{N\times N}=\widetilde{C}_\chi+\widetilde{C}_\varepsilon.
\end{equation}


\section{Large sample theory}\label{sec4}
\renewcommand{\theequation}{4.\arabic{equation}}
\setcounter{equation}{0}

In this section, we first give some regularity conditions and then present the convergence properties for the estimates developed in Section \ref{sec3}.

\subsection{Regularity conditions}\label{sec4.1}

\begin{assumption}\label{ass:1}

{\em (i) Let $\{G_t\}$ be a stationary sequence of $q$-dimensional random vectors with mean zero. There exists a positive definite matrix $\Sigma_G={\sf E}\left[G_tG_t^{^\intercal}\right]$ such that
\[\left\Vert \frac{1}{T}\sum_{t=1}^TG_tG_t^{^\intercal}-\Sigma_G\right\Vert=o_P(1).\]
}

{\em (ii) There exists a positive definite matrix $\Sigma_\Lambda$ such that
\[
\left\Vert \frac{1}{N}\sum_{i=1}^N \int_{u\in{\mathbb C}_i}\Lambda_i(u)\Lambda_i(u)^{^\intercal}du-\Sigma_\Lambda\right\Vert=o(1).
\]
The factor loading operator ${\mathbf B}_{ik}$ satisfies that $\max\limits_{1\leq i\leq N}\max\limits_{1\leq k\leq k_\ast}\Vert {\mathbf B}_{ik}\Vert_{\rm O}\leq m_B$, where $m_B$ is a positive constant.}

{\em (iii) The sieve approximation error $\{\eta_{tk}\}$ in (\ref{eq2.6}) is stationary (over $t$), satisfying that $\max\limits_{1\leq k\leq k_\ast}{\sf E}\left[\Vert\eta_{kt}\Vert_2^2\right]=O\left(\xi_q^2\right)$, where $\xi_q\rightarrow0$ as $q\rightarrow\infty$.}

\end{assumption}

\begin{assumption}\label{ass:2}

{\em (i) Let $\{\varepsilon_{t}\}$ be a stationary sequence of zero-mean ${\mathscr H}_X$-valued random elements, independent of $\{G_t\}$ and $\{\eta_{tk}\}$. There exists a positive constant $m_\varepsilon$ such that
\[
\sum_{s=1}^T\sum_{i=1}^N \left\vert{\sf E}\left[\langle \varepsilon_{it},\varepsilon_{is}\rangle\right]\right\vert\leq m_\varepsilon N,\ \ \forall\ \ 1\leq t\leq T,
\]
and
\[{\sf E}\left(\sum_{i=1}^N \left(\langle \varepsilon_{it},\varepsilon_{is}\rangle-{\sf E}\left[\langle \varepsilon_{it},\varepsilon_{is}\rangle\right]\right)\right)^2\leq m_\varepsilon N,\ \ \forall\ 1\leq s,t\leq T.
\]

}

{\em (ii) For $g_i\in{\mathscr H}_i$, any deterministic function defined on ${\mathbb C}_i$, we have
\[{\sf E}\left(\sum_{i=1}^N\langle g_i,\varepsilon_{it}\rangle\right)^2 \propto \sum_{i=1}^N \Vert g_i\Vert_2^2.
\]
}

\end{assumption}

\begin{assumption}\label{ass:3}

{\em (i) Let $N,T,q$ and $\xi_q$ satisfy
\[q^{2}\left(T^{-1/2}+qN^{-1/2}+q\xi_q\right)\rightarrow0
\]
and
\[\left\{[\log(N\vee T)]^{3+1/\gamma}+q^2[\log(N\vee T)]^{1+1/\gamma}\right\}T^{-1}=o(1),\]
where $\gamma$ is defined in Assumption \ref{ass:3}(ii) below.
}

{\em (ii) For any $k=1,\cdots,q$ and $i,j=1,\cdots,N$, the joint processes $\{(G_{tk},\varepsilon_{it}): t=1,2,\cdots\}$ and $\{(\varepsilon_{it},\varepsilon_{jt}): t=1,2,\cdots\}$ are stationary and $\tau$-mixing with $\tau(n)\leq\theta_1\exp\left\{-(\theta_2n)^\gamma\right\}$, where $\theta_1,\theta_2,\gamma>0$, and $G_{tk}$ denotes the $k$-th element of $G_t$.}

{\em (iii) There exist positive constants $\nu_0$ and $m_\ast$ such that
\[
\max_{1\leq k\leq q}{\sf E}\left[\exp\left\{\nu_0|G_{tk}|^2\right\}\right]+\max_{1\leq i\leq N}{\sf E}\left[\exp\left\{\nu_0\|\varepsilon_{it}\|_2^2\right\}\right]\leq m_\ast.
\]
}

\end{assumption}

\begin{remark}\label{re:1}

Assumption \ref{ass:1} imposes some fundamental conditions on $G_t$ and $\Lambda_i(\cdot)$, which are similar to the assumptions in \cite{BN02}, \cite{FLM13} and \cite{TNH23b}. The high-level convergence condition in Assumption \ref{ass:1}(i) is essentially a weak law of large numbers for $G_tG_t^{^\intercal}$ (with diverging size). Assumption \ref{ass:1}(ii) indicates that the $q$-dimensional real-valued factors $G_t$ are pervasive. We conjecture that the methodology and theory (with modified convergence rates) may remain valid when some factors are weak. The uniform boundedness condition on the factor loading operators is not uncommon in the literature. For instance, the factor loading vectors are often assumed to be bounded in classic factor models \citep[e.g., Assumption 4 in][]{FLM13}. Assumption \ref{ass:1}(iii) implies that the sieve approximation error converges to zero at the $\xi_q$-rate, which, together with Assumption \ref{ass:1}(ii), indicates that $\chi_{it}^\eta$ in (\ref{eq2.7}) also converges at the $\xi_q$-rate.

Assumption \ref{ass:2} contains some high-level moment conditions on the functional idiosyncratic components $\varepsilon_{it}$, indicating that $\varepsilon_{it}$ are allowed to be weakly correlated over $i$ and $t$. They are similar to the conditions used by \cite{BN02} and \cite{FLM13} on the real-valued idiosyncratic components. It is straightforward to verify Assumption \ref{ass:2} when $\varepsilon_{it}$ are independent over $i$ and $t$.

Assumption \ref{ass:3}(i) imposes some mild restrictions on $q$, $N$ and $T$. The dimension $q$ may diverge at a slow polynomial rate of $N\wedge T$, whereas $N$ can be ultra-large, diverging at an exponential rate of $T$. The $\tau$-mixing dependence on the stationary processes is introduced by \cite{DDLLLP07} and \cite{W10} for real-valued random variables and further extended by \cite{BZ19} to Banach-valued random elements. Appendix C gives the concept of $\tau$-mixing dependence.  We refer to \cite{BZ19} for some examples (such as the functional AR(1) process) satisfying the $\tau$-mixing dependence. Furthermore, with the sub-Gaussian moment condition in Assumption \ref{ass:3}(iii), we may adopt \cite{BZ19}'s concentration inequality to derive uniform convergence properties in the high-dimensional setting. It is worth pointing out that our sub-Gaussian condition on $G_{tk}$ and $\varepsilon_{it}$ is weaker than the uniform boundedness restriction in Assumption (H1) of \cite{TNH23b}. This relaxation is due to the truncation technique used in our mathematical proofs. A similar sub-Gaussian condition is also assumed by \cite{LQW23}.

\end{remark}


\subsection{Convergence properties}\label{sec4.2}

We derive the convergence properties in the so-called large panel setting, i.e., when $N$ and $T$ diverge to infinity jointly. We start with the convergence property for the PCA estimators of factors and functional factor loadings. Let $V$ be a $q\times q$ diagonal matrix with the diagonal elements being the first $q$ largest eigenvalues of $\frac{1}{T}\Delta$ (arranged in the decreasing order), and define the following $q\times q$ rotation matrix:
\[R=V^{-1}\left(\frac{1}{T}\widetilde G^{^\intercal}G\right)\left[\frac{1}{N}\sum_{i=1}^N \int_{u\in{\mathbb C}_i}\Lambda_i(u)\Lambda_i(u)^{^\intercal}du\right],\]
where $G=\left(G_1,\cdots,G_T\right)^{^\intercal}$ and $\widetilde G=\left(\widetilde G_1,\cdots,\widetilde G_T\right)^{^\intercal}$. We may show that this random rotation matrix $R$ is asymptotically invertible, see Lemma \ref{le:B.2}.

\renewcommand{\theprop}{4.\arabic{prop}}
\setcounter{prop}{0}

\begin{prop}\label{prop:4.1}

{\em Suppose that Assumptions \ref{ass:1}, \ref{ass:2} and \ref{ass:3}(i) are satisfied.

(i) For the PCA estimator of $G_t$, we have the following mean square convergence:
\begin{equation}\label{eq4.1}
\frac{1}{T}\sum_{t=1}^T\left\vert \widetilde G_t-R G_t\right\vert_2^2=O_P\left(q\left(T^{-1}+q^2N^{-1}+q^2\xi_q^2\right)\right),
\end{equation}
where $\xi_q$ is defined in Assumption \ref{ass:1}(iii).

(ii) If, in addition, Assumption \ref{ass:3}(ii)(iii) is satisfied, for the functional factor loading estimate $\widetilde\Lambda_i$, we have the following uniform convergence:
\begin{equation}\label{eq4.2}
\max_{1\leq i\leq N}\Big{|}\Big{|}\Big{|} \widetilde\Lambda_i- (R^{-1})^{^\intercal}\Lambda_i \Big{|}\Big{|}\Big{|}_2=O_P\left(q^{1/2}\left(\left\{q+[\log (N\vee T)]^{1/2+1/(2\gamma)}\right\}T^{-1/2}+q^2N^{-1/2}+q^2\xi_q\right)\right),
\end{equation}
where $\gamma$ is defined in Assumption \ref{ass:3}(ii).
}

\end{prop}

\begin{remark}\label{re:2}

The mean square convergence rate in (\ref{eq4.1}) is slower than the rates obtained by \cite{BN02} and \cite{TNH23b}. This is due to a diverging number of factors and the existence of sieve approximation error in (\ref{eq2.6}). Similarly, the uniform convergence rate in (\ref{eq4.2}) is also slower than some typical convergence rates in the literature. If we additionally assume that $q$ is a fixed positive integer and $\eta_{tk}\equiv0$, the rates in (\ref{eq4.1}) and (\ref{eq4.2}) can be simplified to
\[
O_P\left(T^{-1}+N^{-1}\right)\ \ \ {\rm and}\ \ \ O_P\left([\log (N\vee T)]^{1/2+1/(2\gamma)}T^{-1/2}+N^{-1/2}\right),
\]
respectively, and the involved rates $N^{-1}$ and $N^{-1/2}$ may disappear when $N\gg T$. It is worth pointing out that the $\tau$-mixing dependence restriction is not required to prove Proposition \ref{prop:4.1}(i).

\end{remark}

The following theorem gives the convergence rates for the estimated covariance functions for the functional idiosyncratic components.

\renewcommand{\thetheorem}{4.\arabic{theorem}}
\setcounter{theorem}{1}

\begin{theorem}\label{thm:4.2}

Suppose that Assumptions \ref{ass:1}--\ref{ass:3} are satisfied, and set the shrinkage parameter $\rho$ as $m_\rho\delta_{N,T,q}$ with $m_\rho$ being a sufficiently large positive constant and
\[
\delta_{N,T,q}=q\left(\left\{q+[\log (N\vee T)]^{1/2+1/(2\gamma)}\right\}T^{-1/2}+q^2N^{-1/2}+q^2\xi_q\right).
\]
Then we have
\begin{equation}\label{eq4.3}
\max_{1\leq i\leq N}\max_{1\leq j\leq N}\left\Vert \widetilde{C}_{\varepsilon,ij}-C_{\varepsilon,ij}\right\Vert_{\rm S}=O_P\left(\delta_{N,T,q}\right),
\end{equation}
and
\begin{equation}\label{eq4.4}
\max_{1\leq i\leq N}\sum_{j=1}^N\left\Vert \widetilde{C}_{\varepsilon,ij}-C_{\varepsilon,ij}\right\Vert_{\rm S}=O_P\left(\varpi_N\delta_{N,T,q}^{1-\iota}\right),
\end{equation}
where $\varpi_N$ and $\iota$ are defined in (\ref{eq3.3}).

\end{theorem}

\begin{remark}\label{re:3}

The uniform convergence results in (\ref{eq4.3}) and (\ref{eq4.4}) can be seen as a natural functional extension of the large covariance matrix estimation theory in the matrix maximum and $\ell_1$ norms, respectively \citep[e.g.][]{BL08, FLM13}. The uniform convergence rates are comparable to those rates derived in the literature \citep[e.g., Remark 3(b) in][]{FLM13}. As discussed in Remark~\ref{re:2}, the divergence rate of $q$ and the existence of series approximation error slow down our uniform convergence rates. Note that $q^2N^{-1/2}$ may disappear if $N\gg T$, in which case the large dimension $N$ affects the convergence rate in (\ref{eq4.4}) via $\varpi_N$ and $\log(N\vee T)$. Furthermore, if $q$ is fixed, $\eta_{tk}\equiv0$ and $N\gg T$, we may show that $\delta_{N,T,q}$ can be simplified to
\[
\delta_{N,T}=[\log (N\vee T)]^{1/2+1/(2\gamma)}T^{-1/2},
\]
and the uniform convergence rate $O_P\left(\varpi_N\delta_{N,T}^{1-\iota}\right)$ would be similar to the rates in Theorem 1 of \cite{FGQ23} and Theorem 4 of \cite{LQW23} when $\gamma$ is large.

\end{remark}

The following theorem states the uniform convergence property for $\widetilde{C}_{X}$ defined in (\ref{eq3.9}).

\begin{theorem}\label{thm:4.3}

Suppose that the assumptions of Theorem \ref{thm:4.2} are satisfied and in addition $\Sigma_G=I_q$. Then we have
\begin{equation}\label{eq4.5}
\max_{1\leq i\leq N}\max_{1\leq j\leq N}\left\Vert \widetilde{C}_{X,ij}-C_{X,ij}^\ast\right\Vert_{\rm S}=O_P\left(q^{1/2}\delta_{N,T,q}\right),
\end{equation}
where $C_{X,ij}^\ast$ is defined in (\ref{eq3.4}), and furthermore,
\begin{equation}\label{eq4.6}
\max_{1\leq i\leq N}\max_{1\leq j\leq N}\left\Vert \widetilde{C}_{X,ij}-C_{X,ij}\right\Vert_{\rm S}=O_P\left(q^{1/2}\delta_{N,T,q}\right).
\end{equation}

\end{theorem}

\begin{remark}\label{re:4}

Theorem \ref{thm:4.3} shows that $\widetilde{C}_{X,ij}$ uniformly converges to either $C_{X,ij}^\ast$ or $C_{X,ij}$ in the Hilbert-Schmidt norm. This is due to the fact that
\begin{equation}\label{eq4.7}
\max_{1\leq i\leq N}\max_{1\leq j\leq N}\left\Vert C_{\varepsilon,ij}^\ast-C_{\varepsilon,ij}\right\Vert_{\rm S}=O_P\left(\xi_q\right)
\end{equation}
or
\begin{equation}\label{eq4.8}
\max_{1\leq i\leq N}\max_{1\leq j\leq N}\left\Vert C_{\chi,ij}^\ast-C_{\chi,ij}\right\Vert_{\rm S}=O_P\left(\xi_q\right),
\end{equation}
where $C_{\varepsilon,ij}^\ast$ is the covariance function between $\varepsilon_{it}^\ast$ and $\varepsilon_{jt}^\ast$ whereas $C_{\chi,ij}^\ast$ is the covariance function between $\chi_{it}^\ast$ and $\chi_{jt}^\ast$. The uniform approximation in (\ref{eq4.7}) and (\ref{eq4.8}) can be verified using Assumption \ref{ass:1}(iii) and Lemma \ref{le:B.3}. When $q$ is fixed, the uniform convergence rates in (\ref{eq4.5}) and (\ref{eq4.6}) would be the same as that in (\ref{eq4.3}).

\end{remark}

Due to the low-rank plus sparse functional matrix structure for $C_X^\ast$ and $C_X$, we cannot derive the uniform convergence property in the functional version of matrix $\ell_1$ norm as in (\ref{eq4.4}). To address this problem, \cite{FLM13} recommends measuring the relative error of large low-rank plus sparse covariance matrix estimation for a high-dimensional random vector. However, it seems difficult to directly extend their relative error measurement to the setting of high-dimensional functional data due to the following reasons. First, the inverse of the large covariance matrix is often involved in defining the relative error measurement, but it seems difficult to compute the inverse of the large matrix of covariance operators ${\mathbf C}_X$. Second, although the uniform approximation properties (\ref{eq4.7}) and (\ref{eq4.8}) hold in the functional version of the matrix maximum norm, it is non-trivial to derive similar approximation properties under the relative error measurement. Although it is difficult to provide a theoretical justification via the relative error measurement, in the simulation study we will report some results on two types of its discrete approximation, illustrating the finite-sample performance of $\widetilde{C}_{X}$.


\section{Numerical studies}\label{sec5}
\renewcommand{\theequation}{5.\arabic{equation}}
\setcounter{equation}{0}

In this section, we first introduce an easy-to-implement criterion to consistently estimate the factor number $q$ and discuss selection of the tuning parameter $\rho$ in the functional shrinkage. Then, we present Monte-Carlo simulation and empirical studies.

\subsection{Practical issues in the estimation procedure}\label{sec5.1}

The functional PCA algorithm proposed in Section \ref{sec3} requires accurate estimation of $q$, the dimension of latent $G_t$. There have been extensive studies on the factor number selection in the context of high-dimensional real-valued time series. For example, \cite{BN02} introduces some information criteria to consistently estimate the factor number, and \cite{LY12} and \cite{AH13} propose a simple ratio criterion by comparing ratios of the estimated eigenvalues. The factor number is often assumed to be fixed and does not change with the size of data in the aforementioned literature. In this paper, we introduce a modified information criterion to consistently estimate the factor number $q$, which can diverge to infinity slowly\footnote{\cite{LLS17} consider a similar scenario in the factor number selection for real-valued panel time series.}. Let $\nu_k(\Delta/T)$ be the $k$\textsuperscript{th} largest eigenvalue of $\Delta/T$ with $\Delta$ defined in (\ref{eq3.5}), and define
\begin{equation}\label{eq5.1}
\widetilde q=\argmin_{1\leq k\leq q_{\max}}\left\{\nu_k(\Delta/T)+k\cdot \phi_{N,T}\right\}-1,
\end{equation}
where $\phi_{N,T}$ in the penalty parameter and $q_{\max}$ is a user-specified positive integer. A similar criterion is also adopted by \cite{AX17} to determine the number of high-frequency latent factors. The selection criterion in (\ref{eq5.1}) amends \cite{BN02}'s information criterion which replaces $\nu_k(\Delta/T)$ by summation of $\nu_j(\Delta/T)$ over $j>k$ and does not require ``-1" adjustment. It is worth pointing out that the amended information criterion~\eqref{eq5.1} is easier to implement and its theoretical justification is straightforward. Proposition \ref{prop:5.1} below derives its consistency property.

\renewcommand{\theprop}{5.\arabic{prop}}
\setcounter{prop}{0}

\begin{prop}\label{prop:5.1}

{\em Suppose that Assumptions \ref{ass:1}, \ref{ass:2} and \ref{ass:3}(i) hold, and $q$, the number of real-valued factors, is positive and may diverge slowly to infinity,  
\begin{equation}\label{eq5.2}
\phi_{N,T}\rightarrow0,\quad q\left(N^{-1/2}+\xi_q\right)+T^{-1/2}=o(\phi_{N,T}).
\end{equation}
Then we have ${\sf P}\left(\widetilde{q}= q\right)\rightarrow1$. }

\end{prop}

The shrinkage estimation of $C_\varepsilon$ is often sensitive to the choice of the tuning parameter $\rho$. \cite{BL08} recommend the cross-validation method to select $\rho$ when the observations are independent and real-valued, see also \cite{FGQ23} for the extension to functional-valued observations. However, since the functional time series observations are serially correlated over time, satisfying the functional factor model structure, the cross-validation method may no longer work well in our setting. We modify the traditional cross-validation as follows.

\begin{description}

\item \textsc{Step 1}:\  Let $\widetilde\varepsilon_t=\left(\widetilde{\varepsilon}_{1t},\cdots,\widetilde\varepsilon_{Nt}\right)^{^\intercal}$ with $\widetilde\varepsilon_{it}=X_{it}-\widetilde\Lambda_{i}^{^\intercal}\widetilde G_t$. Use a rolling window of
size $\lfloor T/2\rfloor+K$ and divide the estimated functional idiosyncratic components $\widetilde\varepsilon_t$ within each window into two sub-samples of sizes $T_{1}=\left\lfloor \frac{T}{2}\big(1-\frac{1}{\log
(T/2)}\big)\right\rfloor $ and $T_{2}=\lfloor T/2\rfloor-T_{1}$ by leaving out
K observations in-between, where $\lfloor\cdot\rfloor$ denotes the floor function.

\item \textsc{Step 2}:\ For the $k$-th rolling window, we compute the shrinkage estimate $\widetilde{C}_{\varepsilon,\rho}^{(k)}=(\widetilde{C}_{\varepsilon,\rho,ij}^{(k)})_{N\times N}$ as in (\ref{eq3.8}) using the first sub-sample, where we make its dependence on $\rho$ explicitly, and the conventional estimate $\widehat{C}_{\varepsilon}^{(k)}=(\widehat{C}_{\varepsilon,ij}^{(k)})_{N\times N}$ (without shrinkage) as in (\ref{eq3.7}) using the second sub-sample, $k=1,\cdots,K_\circ$ with $K_\circ=\lfloor T/(2K)\rfloor$. Determine the shrinkage parameter $\rho$ by minimising
\begin{equation}
\frac{1}{N^2K_\circ}\sum\limits_{k=1}^{K_\circ}\sum_{i=1}^N\sum_{j=1}^N\left\|  \widetilde{C}_{\varepsilon,\rho,ij}^{(k)}-\widehat{C}_{\varepsilon,ij}^{(k)}\right\|_{\rm S}^{2}.\label{eq:CV}
\end{equation}

\end{description}

The above selection criterion is introduced by \cite{CLL19} for large covariance matrix estimation of weakly dependent real-valued time series. The reason for leaving out $K$ observations between the two subsamples in each rolling window is to make these two subsamples have negligible correlation. In practice, for weakly dependent functional time series, we may set $K=10$.

\subsection{Simulation study}\label{sec:5.2}

We start with the description of the data generating process. For simplicity, let $k_\ast=1$ in the fully functional factor model (\ref{eq2.1}) or (\ref{eq2.2}). The functional factor process $\{F_t: t=1,2,\cdots\}$ is defined by (\ref{eq2.6}), i.e.,
\[
F_{t}(u)=\Phi(u)^{^\intercal}G_t+\frac{1}{q}\eta_t^\ast(u),\ \ u\in{\mathbb C}=[0,1],
\]
where $\eta_{t}^\ast(\cdot)$ is generated from a Brownian bridge, $\Phi(\cdot)$ is a $q$-dimensional vector of Fourier basis functions, $G_t$ follows a VAR$(1)$ model:
\[
G_t = {\mathbf A}G_{t-1}+\zeta_t
\]
with ${\mathbf A}$ being a $q\times q$ coefficient matrix with its element $A_{jk} = 0.25^{|j-k|+1}$ for $j,k = 1,\dots,q$ and $\zeta_t$ being independently generated by a $q$-dimensional standard normal distribution. Each function is generated on a common set of 21 equally-spaced grid points on $[0,1]$. The number of real-valued factors, $q$, is set as $5,10$ or $15$. We simulate the factor loadings $B_{i}(u,v)$ via
\[
B_{i}(u,v) = \sum^{50}_{j=1}c_{i,j}\phi_j(u)\phi_j(v),
\]
where $c_{i,j}$ is independently generated from a standard normal distribution, and $\{\phi_1,\cdots,\phi_{50}\}$ is a set of Fourier basis functions.

As in \cite{FGQ23}, we simulate the functional idiosyncratic term as
\[
\varepsilon_{it}(u)=\sum_{j=1}^{50}\frac{1}{j}\theta_{it,j}\phi_j(u),\quad i=1,\cdots,N,\quad t=1,\cdots,T,
\]
where $\phi_j$ is the Fourier basis function and $\theta_t=(\theta_{1t}^{^\intercal},\cdots,\theta_{Nt}^{^\intercal})^{^\intercal}$ with $\theta_{it}=(\theta_{it,1},\cdots,\theta_{it,50})^{^\intercal}$ are independently generated from a multivariate Gaussian distribution with mean zero and covariance matrix $\Sigma\in {\mathscr R}^{50N\times 50N}$. Write $\Sigma=(\Sigma_{jk})_{N\times N}$ with $\Sigma_{jk}\in {\mathscr R}^{50\times 50}$ being the $(j,k)$ block, $1\leq j,k\leq N$. The functional sparsity pattern on $C_\varepsilon$ may be characterised by a sparsity structure in $\Sigma$. In the simulation study, we define $\Sigma_{jk} = \omega_{jk}\Omega$ with $\Omega=\text{diag}(1^{-2},\cdots,50^{-2})$ and
\[ \omega_{jk} = \left\{ \begin{array}{ll}
        \left(1-\frac{|j-k|}{10}\right)_{+} & \mbox{for $j,k=1,\cdots,\frac{N}{2}$};\\
       4I(j=k) & \mbox{for $j,k =\frac{N}{2}+1,\cdots,N$},\end{array} \right.
\]
where $I(\cdot)$ denotes the binary indicator function.

We apply the developed estimation method to the simulated data and compute the difference between the true and estimated functional covariance matrices over 200 replications. As in \cite{FGQ23}, we compute the functional version of $\ell_1$ and $\ell_2$ matrix norms for the functional idiosyncratic covariance matrix estimation (see Table~\ref{tab:1}), and the matrix maximum norms (see Table \ref{tab:2}), providing the finite-sample justification for Theorem \ref{thm:4.2}. These functional $\ell_1$, $\ell_2$ and maximum matrix norms are defined as 
\begin{align*}
\ell_1(C_{\varepsilon}) &= \max_{1\leq i\leq N} \sum^{N}_{j=1}\sqrt{\frac{1}{21^2}\sum_{s=1}^{21}\sum_{t=1}^{21}\left\vert \widetilde{C}_{\varepsilon,ij}(u_s,u_t) - C_{\varepsilon,ij}(u_s,u_t) \right\vert^2}, \\
\ell_2(C_{\varepsilon}) &= \sqrt{\frac{1}{21^2}\sum_{i=1}^N\sum^{N}_{j=1}\sum_{s=1}^{21}\sum_{t=1}^{21}\left\vert \widetilde{C}_{\varepsilon,ij}(u_s,u_t) - C_{\varepsilon,ij}(u_s,u_t) \right\vert^2}, \\
\ell_{\max}(C_{\varepsilon}) &= \max_{1\leq i\leq N}\max_{1\leq j\leq N}\sqrt{\frac{1}{21^2}\sum_{s=1}^{21}\sum_{t=1}^{21}\left\vert \widetilde{C}_{\varepsilon,ij}(u_s,u_t) - C_{\varepsilon,ij}(u_s,u_t) \right\vert^2}.
\end{align*}
We consider four shrinkage functions in the functional shrinkage technique: hard thresholding (Hard), soft thresholding (Soft), SCAD and adaptive lasso (Alasso), and include the sample covariance function estimation (sample) as a benchmark. The tuning parameter involved in the functional shrinkage is selected via the modified cross-validation introduced in Section \ref{sec5.1}. Table \ref{tab:1} reports the functional $\ell_1$- and $\ell_2$-norm estimation errors for functional idiosyncratic covariance matrices. Both the $\ell_1$- and $\ell_2$-norm estimation errors decrease significantly when $T$ increases from 100 to 200; the $\ell_1$-norm shrinkage estimation errors slightly increase when $N$ increases from $50$ to 200 but $T$ is fixed; whereas the increase of $\ell_2$-norm estimation errors is more substantial. The estimation performance is generally stable as $q$ increases from~5 to~15. The use of functional shrinkage significantly outperforms the naive sample covariance function without shrinkage. In particular, the adaptive lasso performs best among the four shrinkage functions. Table \ref{tab:2} further compares the functional max-norm estimation errors between the adaptive lasso\footnote{The results are almost the same for the other three shrinkage methods.} and sample covariance function. Unlike Table \ref{tab:1}, their estimation performance is very close.

To measure the covariance matrix estimation accuracy of the observed functional observations simulated by the dual functional factor model, due to the existence of spiked eigenvalues, we cannot adopt the functional $\ell_1$- and $\ell_2$-norm estimation errors both of which are divergent as $N$ increases. As discussed in Remark \ref{re:4}, we define the following two relative-error measurements:
\begin{eqnarray}
{\sf RE}_1&=&\frac{1}{21^2}\sum_{i=1}^{21}\sum_{j=1}^{21}\frac{1}{\sqrt{N}}\left\Vert C_X^{-1/2}(u_i,u_i)\left[\widetilde{C}_X(u_i,u_j)-C_X(u_i,u_j)\right]C_X^{-1/2}(u_j,u_j)\right\Vert_F,\nonumber\\
{\sf RE}_2&=&\frac{1}{21}\sum_{i=1}^{21}\frac{1}{\sqrt{N}}\left\Vert C_X^{-1/2}(u_i,u_i)\left[\widetilde{C}_X(u_i,u_i)-C_X(u_i,u_i)\right]C_X^{-1/2}(u_i,u_i)\right\Vert_F,\nonumber
\end{eqnarray}
whose results are reported in Table \ref{tab:3}. The relative-error measurements increase when $N$ increases, but decrease when $T$ increases from $100$ to $200$. The adoptive lasso again outperforms the other three shrinkage methods, which is consistent with the finding in Table \ref{tab:1}.

We assess the performance of the amended information criterion in Table \ref{tab:4} by reporting percentages of accurately estimating the true number of factors. When $N=100$ and $200$, we achieve perfect accuracy in selecting the factor number; when $N=50$, the proposed information criterion may occasionally over-estimate the factor number, which has negligible impact on the subsequent covariance function estimation.

\newpage

\begin{center}
{\footnotesize\tabcolsep 0.2in
\begin{longtable}{@{}lllrrrrrr@{}}
\caption{The averaged $\ell_1$ and $\ell_2$-norm errors for functional idiosyncratic covariance matrices}\label{tab:1}\\
\toprule
& &  	   &	\multicolumn{3}{c}{$T=100$}   	& \multicolumn{3}{c}{$T=200$} \\
$q$   &  Norm & Shrinkage & $N=50$ & $N=100$ & $N=200$ & $N=50$ & $N=100$ & $N=200$ \\
  \midrule
5 	& $\ell_1(C_{\varepsilon})$ &	Hard & 6.383 & 6.904 & 7.450 & 4.540 & 4.520 & 4.829 \\ 
  & & Soft & 7.045 & 7.582 & 7.991 & 5.754 & 5.976 & 6.500 \\ 
  & & SCAD & 6.932 & 7.518 & 7.954 & 5.248 & 5.517 & 6.154 \\ 
  & & Alasso & 6.442 & 7.010 & 7.511 & 4.657 & 4.688 & 5.079 \\ 
  & & Sample & 26.202 & 52.055 & 104.591 & 19.193 & 37.477 & 75.038 \\ 
\cmidrule{3-9}
    & $\ell_2(C_{\varepsilon})$ & Hard & 7.529 & 11.056 & 16.147 & 5.144 & 7.081 & 10.272 \\ 
  & & Soft & 8.344 & 12.366 & 18.126 & 6.277 & 8.991 & 13.246 \\ 
  & & SCAD & 8.134 & 12.171 & 17.970 & 5.598 & 7.993 & 11.939 \\ 
  & & Alasso & 7.401 & 10.823 & 15.789 & 5.126 & 7.012 & 10.153 \\ 
  & & Sample & 19.883 & 39.288 & 78.504 & 14.696 & 28.635 & 56.887 \\ 
  \midrule
10 	& $\ell_1(C_{\varepsilon})$  &  Hard & 6.476 & 6.913 & 7.380 & 4.220 & 4.382 & 4.618 \\ 
  & & Soft & 7.118 & 7.598 & 7.970 & 5.571 & 5.993 & 6.309 \\ 
  & & SCAD & 7.003 & 7.527 & 7.932 & 4.997 & 5.503 & 5.908 \\ 
  & & Alasso & 6.584 & 7.057 & 7.492 & 4.423 & 4.710 & 4.983 \\ 
  & & Sample & 25.049 & 49.770 & 99.461 & 18.425 & 35.630 & 70.538 \\  
\cmidrule{3-9}
    & $\ell_2(C_{\varepsilon})$ & Hard & 7.782 & 11.256 & 16.357 & 5.126 & 7.079 & 10.242 \\ 
  & & Soft & 8.632 & 12.680 & 18.488 & 6.366 & 9.190 & 13.489 \\ 
  & & SCAD & 8.413 & 12.461 & 18.322 & 5.641 & 8.144 & 12.098 \\ 
  & & Alasso & 7.743 & 11.195 & 16.221 & 5.173 & 7.131 & 10.274 \\ 
  & & Sample & 19.179 & 37.699 & 75.128 & 14.208 & 27.574 & 54.749 \\   
\midrule
15 & $\ell_1(C_{\varepsilon})$ & Hard & 6.634 & 7.110 & 7.604 & 4.564 & 4.472 & 4.723 \\ 
  & & Soft & 7.221 & 7.691 & 8.021 & 5.834 & 6.058 & 6.379 \\ 
  & & SCAD & 7.084 & 7.621 & 7.981 & 5.300 & 5.582 & 5.979 \\ 
  & & Alasso & 6.777 & 7.250 & 7.669 & 4.767 & 4.824 & 5.103 \\ 
  & & Sample & 24.330 & 48.268 & 96.486 & 18.149 & 35.134 & 69.437 \\ 
\cmidrule{3-9}
    & $\ell_2(C_{\varepsilon})$ & Hard & 8.312 & 11.852 & 17.179 & 5.435 & 7.339 & 10.518 \\ 
  & & Soft & 9.112 & 13.197 & 19.120 & 6.659 & 9.476 & 13.832 \\ 
  & & SCAD & 8.871 & 12.964 & 18.917 & 5.947 & 8.431 & 12.430 \\ 
  & & Alasso & 8.356 & 11.878 & 17.151 & 5.492 & 7.443 & 10.659 \\ 
  & & Sample & 18.891 & 36.766 & 73.052 & 14.136 & 27.227 & 53.967 \\  \bottomrule
\end{longtable}}
\end{center}

\newpage

\begin{center}
{\footnotesize\tabcolsep 0.2in
\begin{longtable}{@{}llrrrrrr@{}}
\caption{The averaged max-norm errors for functional idiosyncratic covariance matrices}\label{tab:2}\\
\toprule
&  	   &	\multicolumn{3}{c}{$T=100$}   	& \multicolumn{3}{c}{$T=200$} \\
$q$   & Shrinkage & $N=50$ & $N=100$ & $N=200$ & $N=50$ & $N=100$ & $N=200$ \\
  \midrule
5 &  Alasso & 1.397 & 1.453 & 1.553 & 1.053 & 1.068 & 1.122 \\ 
  & Sample & 1.441 & 1.523 & 1.646 & 1.088 & 1.105 & 1.171 \\  
  \midrule
10 &  Alasso & 1.532 & 1.549 & 1.651 & 1.116 & 1.115 & 1.163 \\ 
   & Sample & 1.542 & 1.574 & 1.689 & 1.131 & 1.138 & 1.190 \\ 
\midrule
15 &  Alasso & 1.684 & 1.720 & 1.795 & 1.207 & 1.191 & 1.222 \\ 
 & Sample & 1.686 & 1.724 & 1.801 & 1.211 & 1.200 & 1.232 \\ 
 \bottomrule
\end{longtable}}
\end{center}

\begin{center}
{\footnotesize\tabcolsep 0.18in
\begin{longtable}{@{}lllrrrrrr@{}}
\caption{The averaged relative errors for functional covariance matrices of the observations}\label{tab:3}\\
\toprule
& & &  	\multicolumn{3}{c}{$T=100$}   	& \multicolumn{3}{c}{$T=200$} \\
$q$   & Norm & Shrinkage & $N=50$ & $N=100$ & $N=200$ & $N=50$ & $N=100$ & $N=200$ \\
\endfirsthead
\toprule
& & & \multicolumn{3}{c}{$T=100$}   	& \multicolumn{3}{c}{$T=200$} \\
$q$   & Norm & Shrinkage & $N=50$ & 100 & 200 & 50 & 100 & 200 \\
\midrule
\endhead
\hline \multicolumn{9}{r}{{Continued on next page}} \\
\endfoot
\endlastfoot
  \midrule	
 5     & RE$_1$ & Hard & 2.162 & 4.283 & 9.218 & 1.685 & 3.453 & 7.390 \\ 
  & & Soft & 2.464 & 5.386 & 13.373 & 1.880 & 4.199 & 10.306 \\ 
  & & SCAD & 2.202 & 5.070 & 13.055 & 1.232 & 2.728 & 6.836 \\ 
  & & Alasso & 1.517 & 3.111 & 7.361 & 1.247 & 2.475 & 5.311 \\ 
\\
    & RE$_2$ & Hard & 3.780 & 7.807 & 17.228 & 2.775 & 5.949 & 13.380 \\ 
  & & Soft & 5.287 & 11.650 & 29.261 & 3.980 & 9.023 & 22.612 \\ 
  & & SCAD & 4.704 & 10.981 & 28.601 & 2.227 & 5.453 & 14.702 \\ 
  & & Alasso & 2.842 & 6.231 & 15.584 & 1.998 & 4.273 & 10.066 \\ 
  \midrule	
10   & $RE_1$   & Hard & 1.382 & 3.004 & 6.758 & 0.977 & 2.178 & 4.847 \\ 
  & & Soft & 1.574 & 3.715 & 9.401 & 1.180 & 2.921 & 7.433 \\ 
  & & SCAD & 1.394 & 3.445 & 9.105 & 0.704 & 1.747 & 4.729 \\ 
  & & Alasso & 0.983 & 2.112 & 5.192 & 0.688 & 1.454 & 3.335 \\ 
\\
    & $RE_2$  & Hard & 2.502 & 5.701 & 13.150 & 1.740 & 4.105 & 9.422 \\ 
  & & Soft & 3.323 & 7.907 & 20.087 & 2.518 & 6.259 & 15.992 \\ 
  & & SCAD & 2.906 & 7.330 & 19.470 & 1.313 & 3.616 & 10.112 \\ 
  & & Alasso & 1.814 & 4.227 & 10.812 & 1.184 & 2.783 & 6.775 \\ 
  \midrule				
15 & $RE_1$   & Hard & 0.986 & 2.305 & 5.372 & 0.687 & 1.630 & 3.751 \\ 
  & & Soft & 1.075 & 2.656 & 6.751 & 0.808 & 2.147 & 5.599 \\ 
  & & SCAD & 0.964 & 2.442 & 6.460 & 0.546 & 1.337 & 3.645 \\ 
  & & Alasso & 0.780 & 1.663 & 4.079 & 0.532 & 1.125 & 2.648 \\ 
\\        
        & $RE_2$    & Hard & 1.698 & 4.337 & 10.467 & 1.181 & 3.063 & 7.315 \\ 
  & & Soft & 2.184 & 5.639 & 14.469 & 1.684 & 4.611 & 12.062 \\ 
  & & SCAD & 1.893 & 5.158 & 13.845 & 0.939 & 2.715 & 7.730 \\ 
  & & Alasso & 1.318 & 3.228 & 8.391 & 0.848 & 2.111 & 5.336 \\ 
 \bottomrule
\end{longtable}}
\end{center}

\begin{table}[!htb]
\centering
\tabcolsep 0.53in
{\footnotesize\caption{The percentages of correctly determining the true factor number $q$.}\label{tab:4}
\begin{tabular}{@{}lllrr@{}}
\toprule
$q$ 	& $T$ & $N=50$ 		 & $N=100$ 	& $N=200$ \\
\midrule
5 	& $T=100$ & 99.5\%          	 & 100\% 		& 100\% \\
	& $T=200$ & 97.5\%  	 & 100\% 		& 100\% \\
\\
10 	& $T=100$ & 100\% 		 & 100\% 	& 100\% \\
	& $T=200$ & 98.5\%         & 100\%	& 100\% \\	
\\
15 	& $T=100$ & 100\% 		 & 100\%	& 100\% \\
	& $T=200$ & 98\% & 100\% 	& 100\% \\	
\bottomrule	
\end{tabular}}
\end{table}

\subsection{Empirical application}\label{sec:5.3}
\begin{figure}[!htb]
\centering
\includegraphics[width=5.9cm]{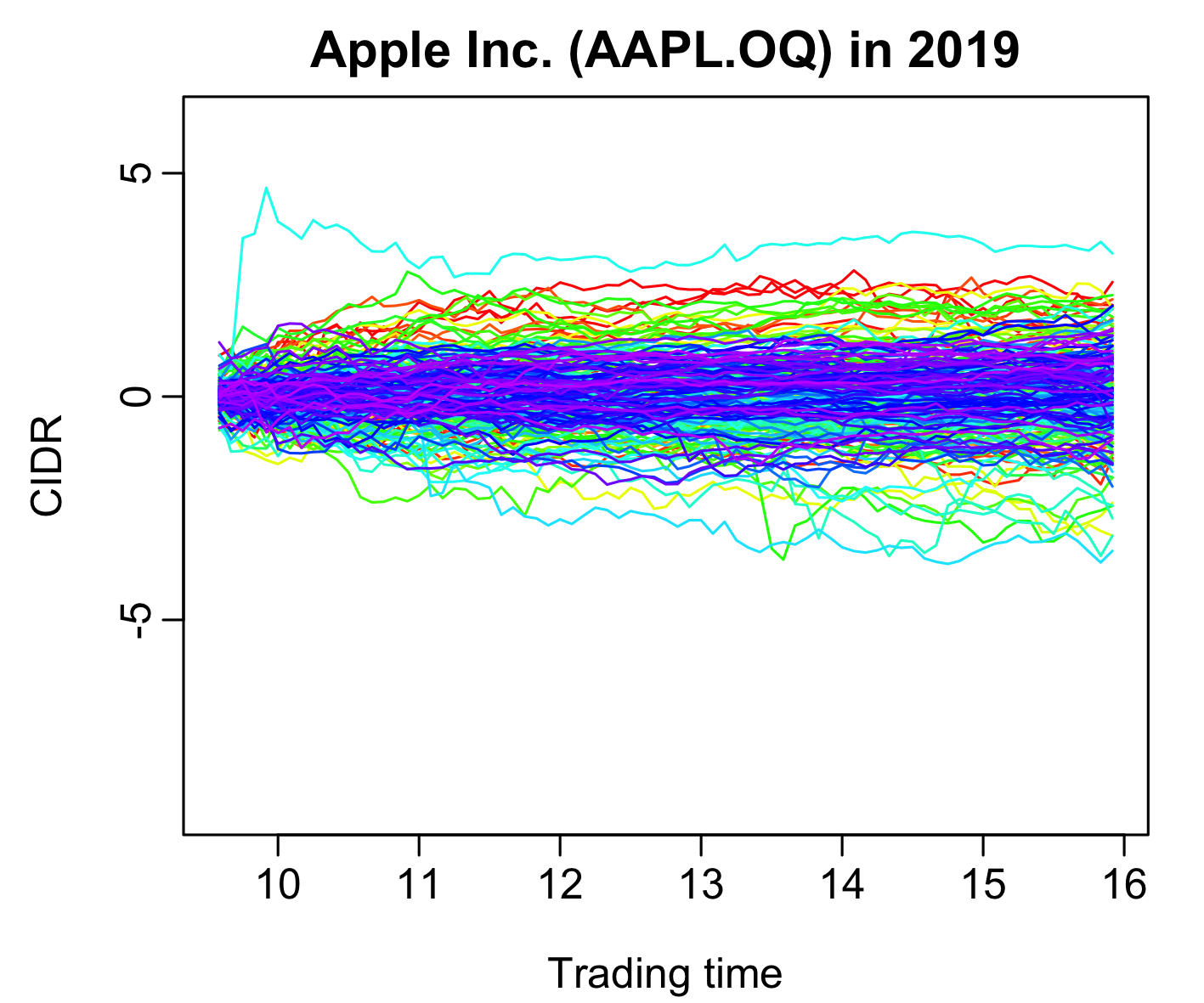}
\includegraphics[width=5.9cm]{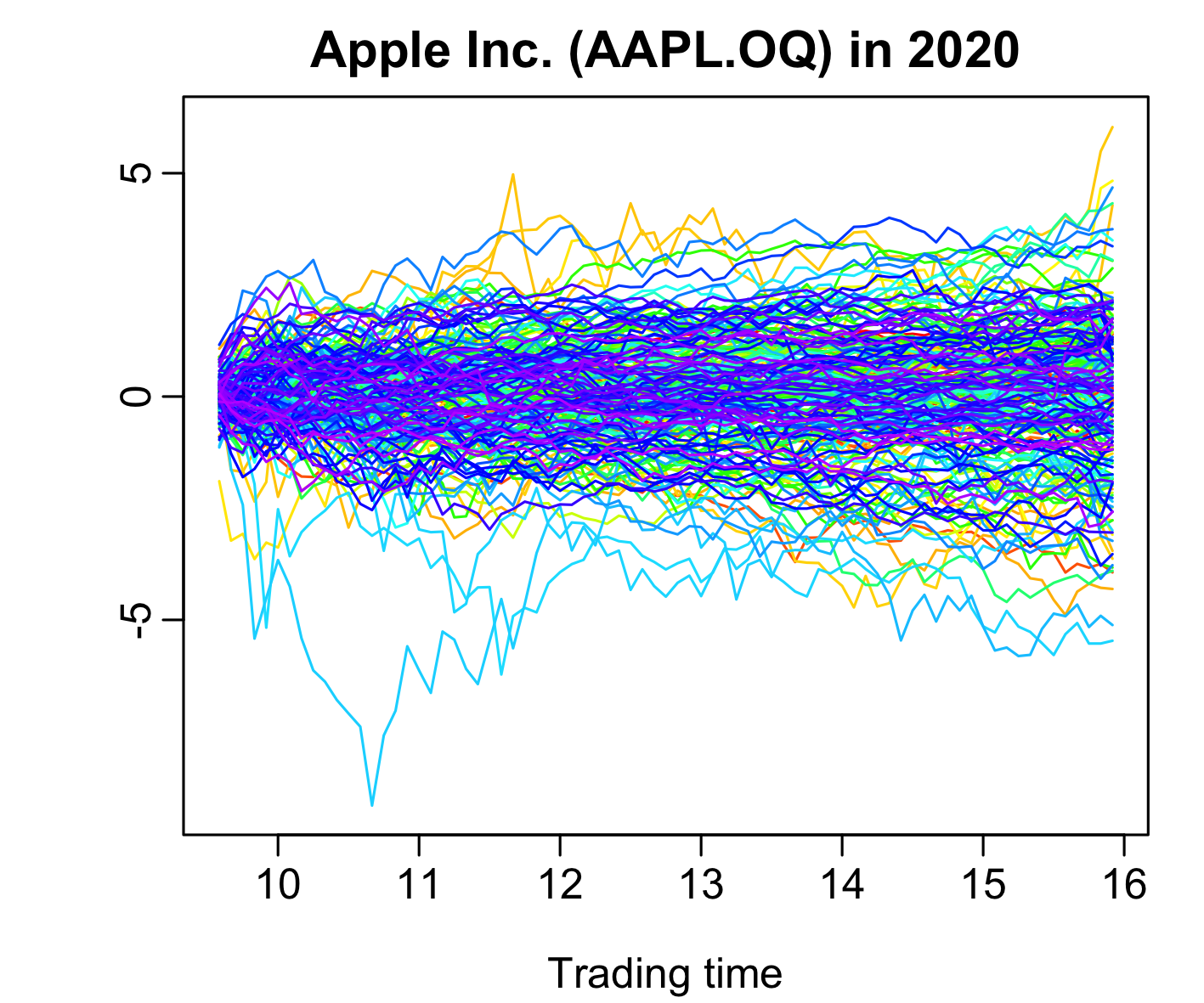}
\includegraphics[width=5.9cm]{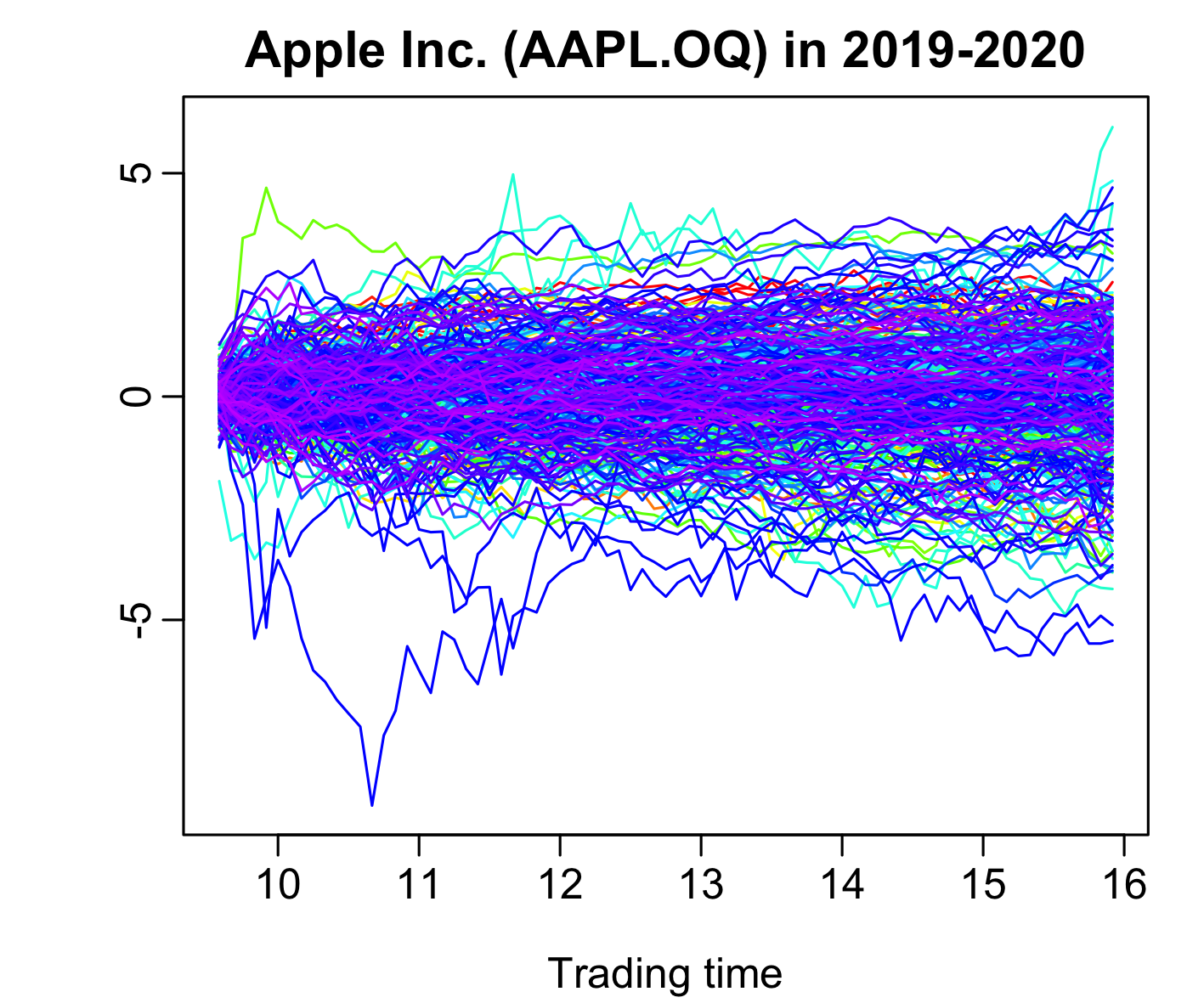}
\caption{Functional time series plot of the CIDRs of Apple Inc. (Tick Symbol AAPL.OQ) in 2019 or/and 2020.}\label{fig:1}
\end{figure}

We apply the developed method to analyse the functional covariance structure of the S\&P 500 data containing $473$ common stocks traded in the years 2019 and 2020, which are available at the Refinitiv Datascope\footnote{\url{https://select.datascope.refinitiv.com/DataScope/}}. We consider their CIDR curves from 2 January 2019 to 31 December 2020. The S\&P 500 index comprises some of the largest companies traded in New York Stock Exchange.  After removing the public holidays and half-day trading days (like Christmas Eve), there are $T_1=242$ trading days in 2019 and $T_2=228$ in 2020. For each trading day, we consider $5$-minute resolution data between 9:30 and 16:00 Eastern Standard Time, and obtain $78$ data points. For asset $i$, let $P_{it}(u_j)$ be the intraday 5-minute close price at time $u_j$ on trading day $t$, and construct a sequence of CIDRs \citep[e.g.,][]{RWZ20}:
\[
X_{it}(u_j) = 100\times [\ln P_{it}(u_j) - \ln P_{it}(u_1)],
\]
where $\ln(\cdot)$ denotes the natural logarithm, $j=2,3,\cdots,78$ and $i=1,2,\cdots,473$. The way CIDR is constructed removes the effect of the starting price. The linear interpolation algorithm in \cite{HAB+23} is adopted to convert discrete data points into a continuous function. In Figure~\ref{fig:1}, we display the CIDRs for Apple Inc., one of the most liquid stocks, in 2019 and 2020.

We fit the  dual functional factor model to the S\&P 500 time series and apply the estimation methodology developed in Section \ref{sec3}. We consider not only the entire time period 2019--2020, but also the two calendar years 2019 and 2020 separately. With the modified information criterion, we select the number of real-valued factors (i.e., $G_t$) to be one for 2019, two for 2020, and two for 2019--2020. As recommended in the simulation, we use the functional shrinkage with adaptive lasso to estimate the functional idiosyncratic covariance matrix. With the amended cross-validation in Section \ref{sec5.1}, we select the tuning parameter as 0.160 for 2019, 0.230 for 2020, and 0.114 for 2019--2020.

\begin{figure}[!htb]
\centering
{\includegraphics[width=5.92cm]{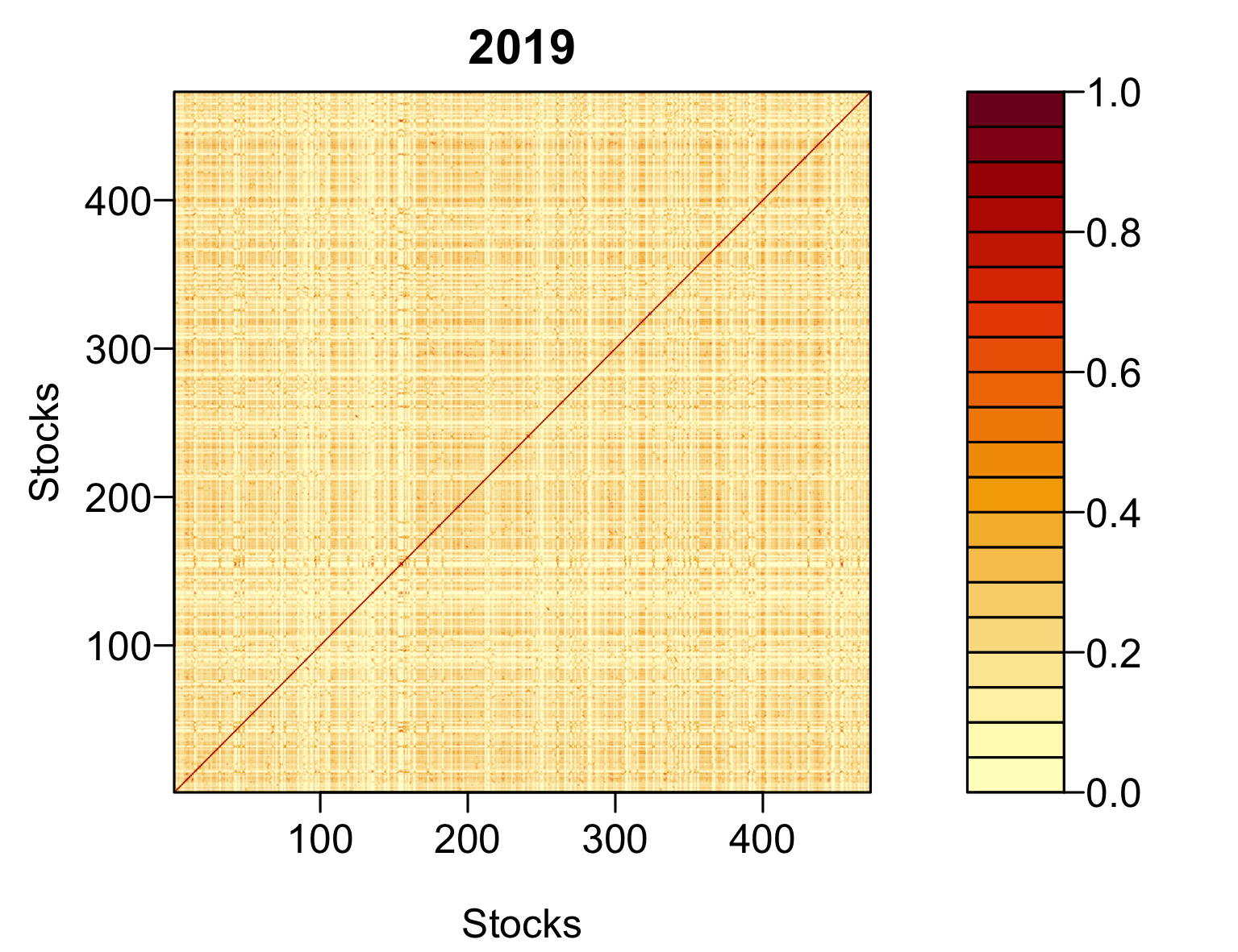}}
{\includegraphics[width=5.92cm]{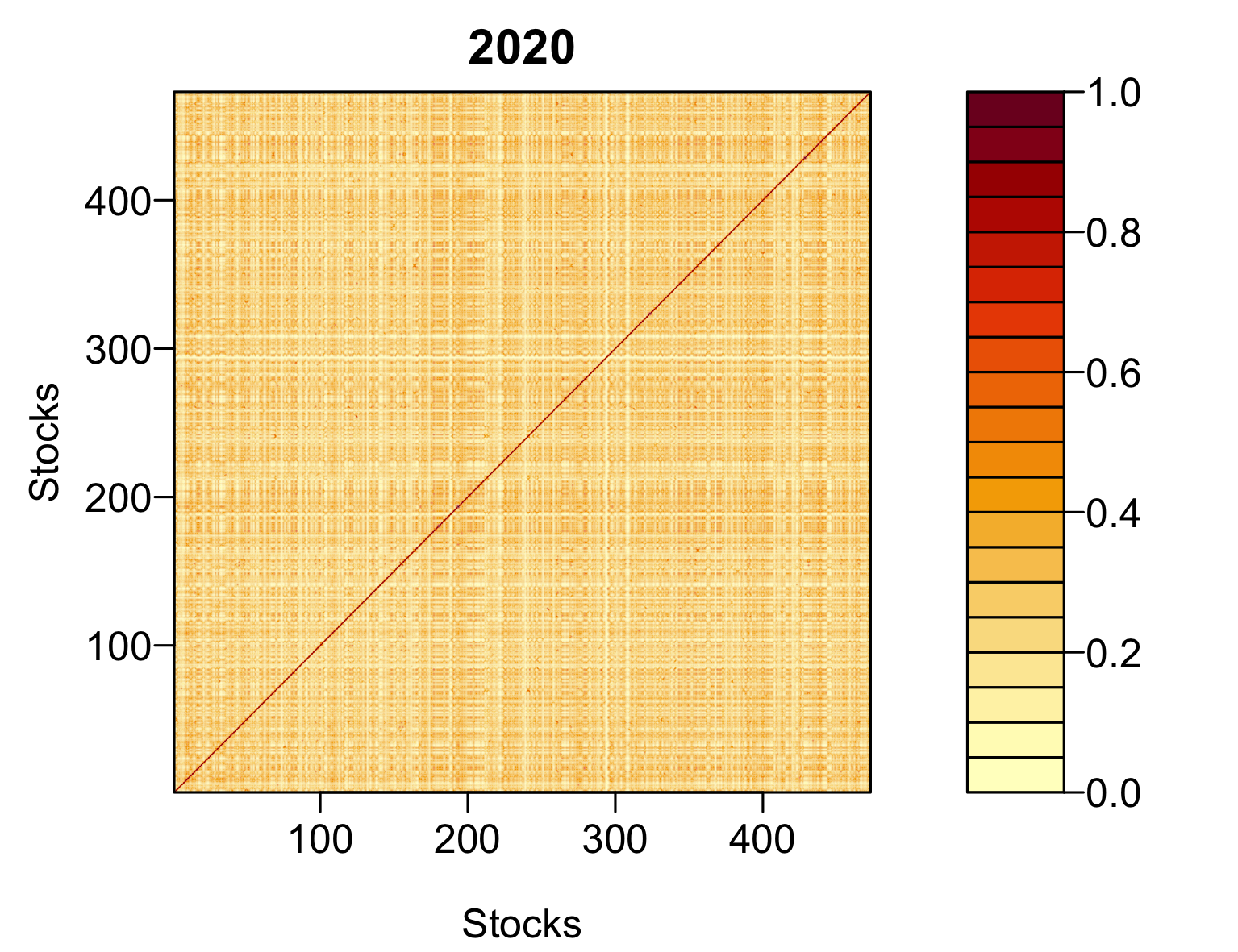}}
{\includegraphics[width=5.92cm]{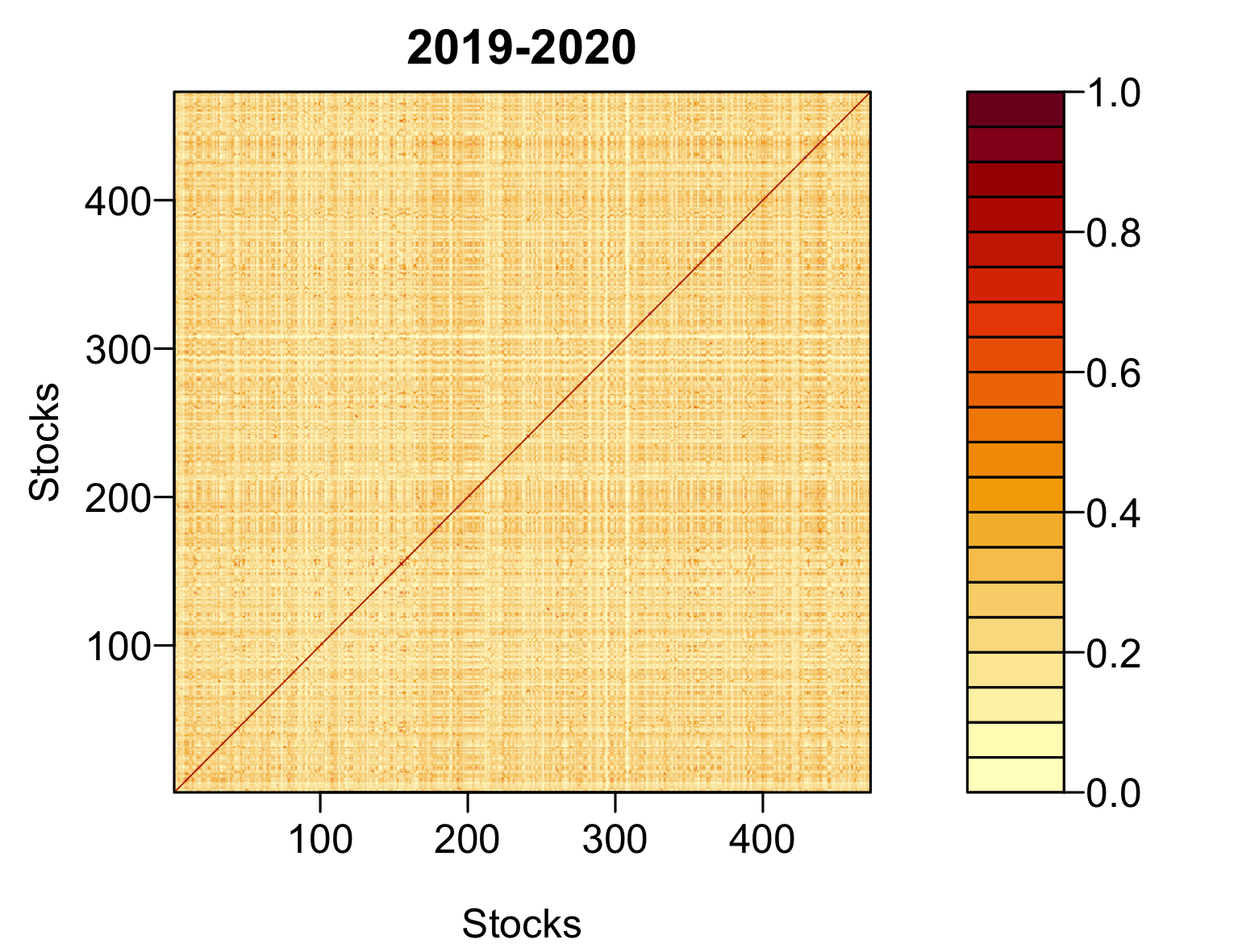}}
\\
{\includegraphics[width=5.92cm]{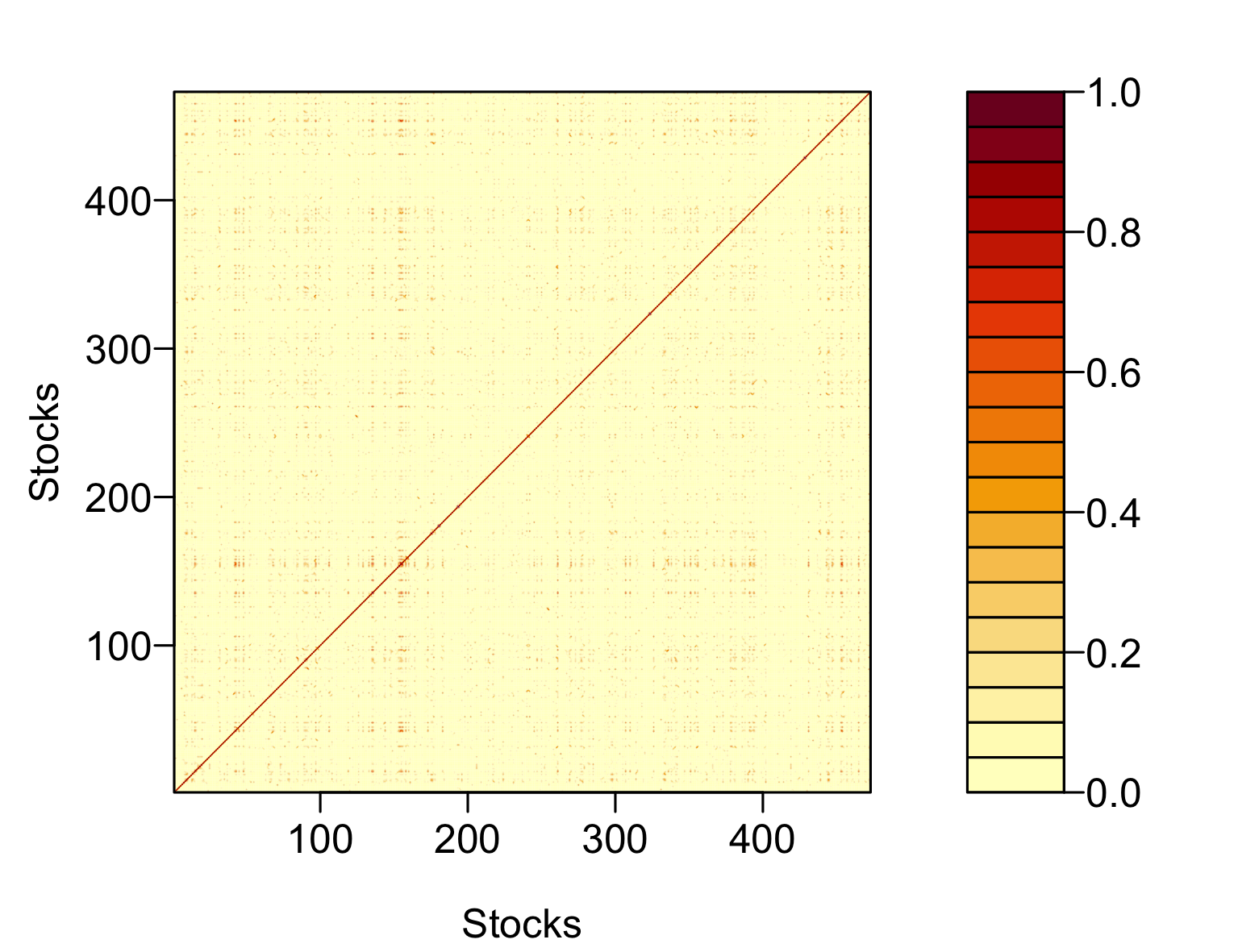}}
{\includegraphics[width=5.92cm]{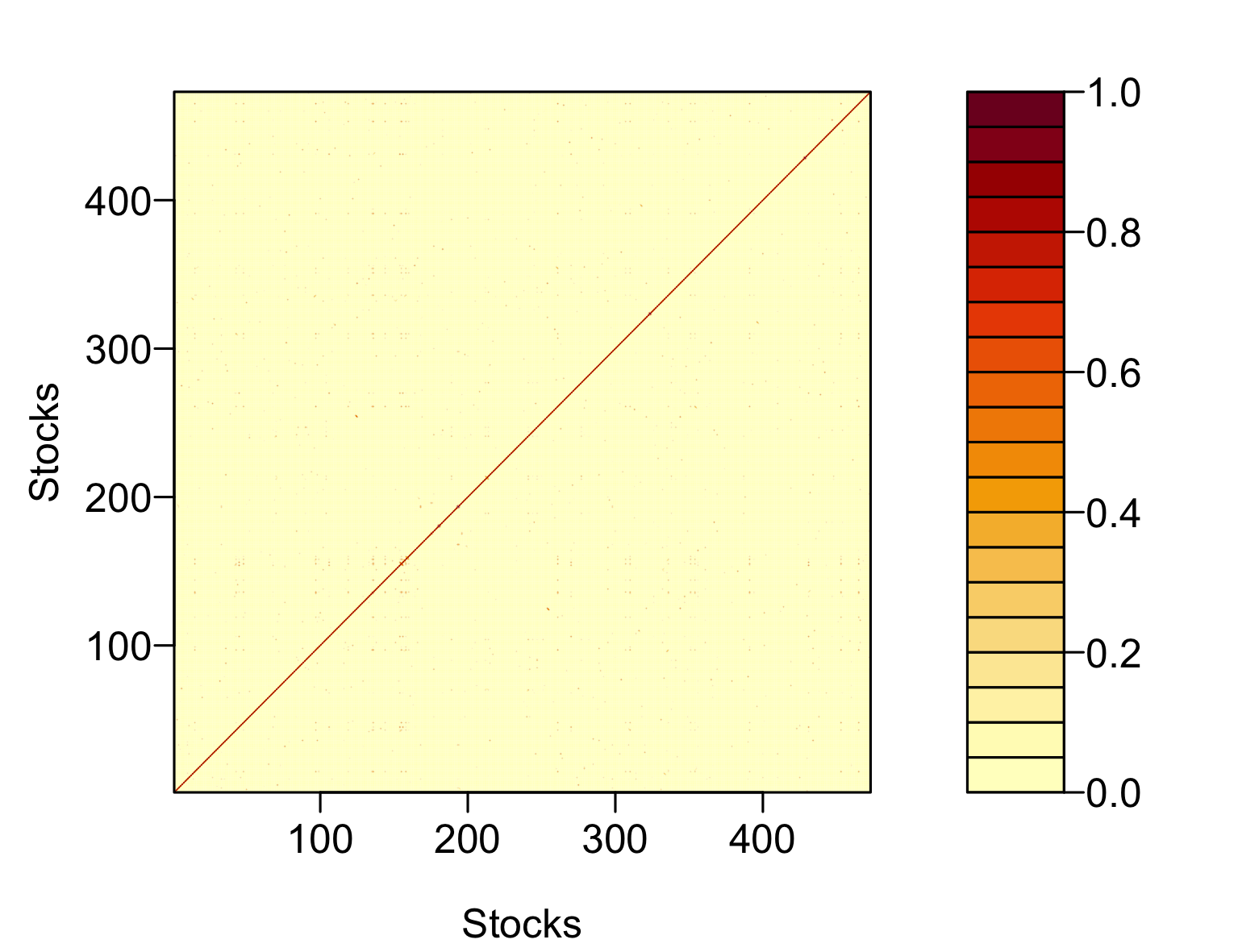}}
{\includegraphics[width=5.92cm]{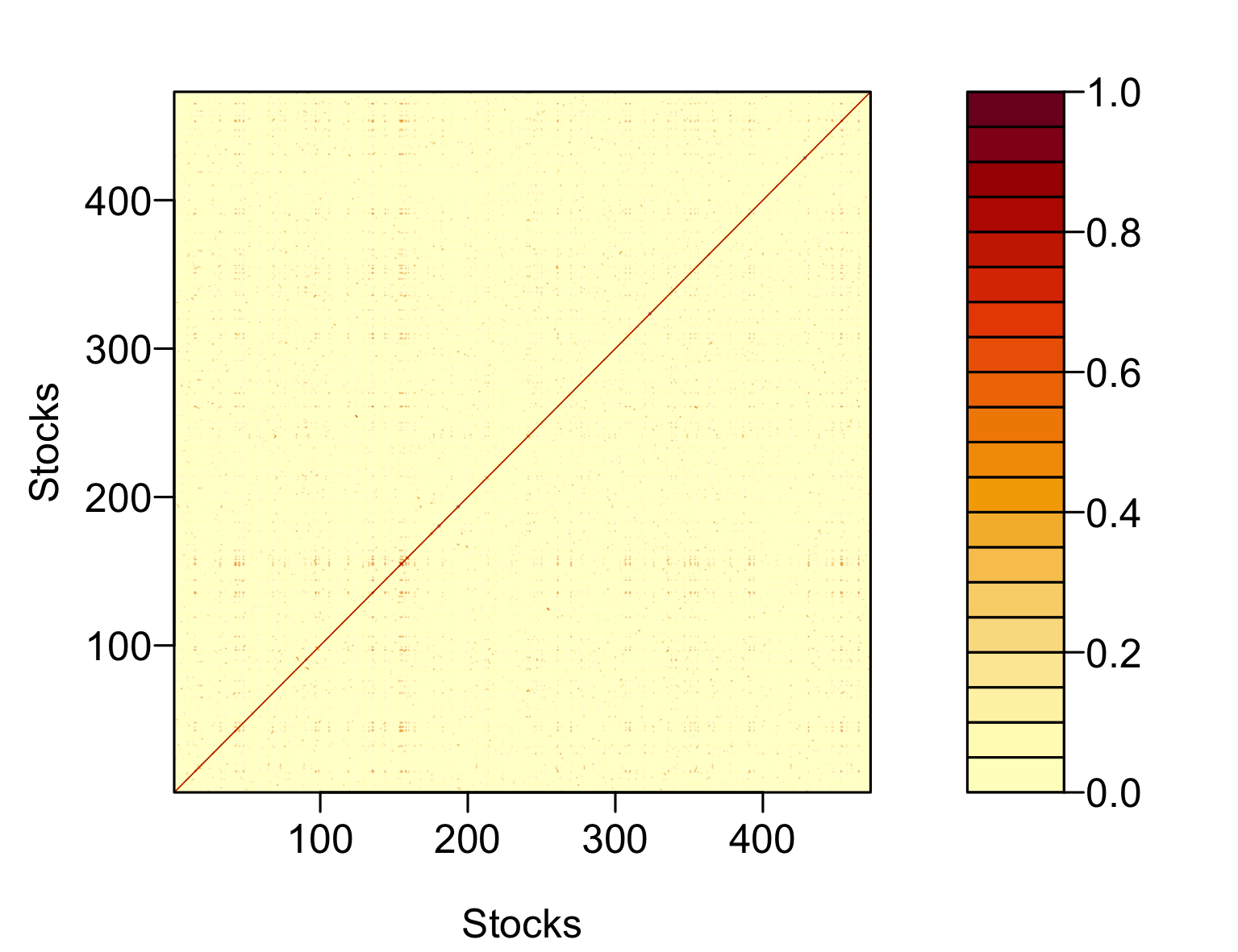}}
\caption{Heat maps of the estimated functional correlation matrices for the observed functional observations (1\textsuperscript{st} row) and idiosyncratic components (2\textsuperscript{nd} row) in 2019 or/and 2020.}\label{fig:2}
\end{figure}

With the estimated covariance functions $\widetilde{C}_{X,ij}$ between stocks, we may further compute the correlation functions whose Hilbert-Schmidt norms are comparable over stocks. Define
\[
\widetilde{\sf cor}(X_i, X_j) = \frac{\widetilde{C}_{X,ij}}{\Vert \widetilde{C}_{X,ii}\Vert_S \Vert\widetilde{C}_{X,jj}\Vert_S},
\]
where $\Vert\cdot\Vert_S$ denotes the Hilbert-Schmidt norm of a covariance function (or operator). The correlation functions between the functional idiosyncratic components are computed in a similar way. The Hilbert-Schmidt norms of the estimated correlation functions are used to plot the heat maps displayed in Figure~\ref{fig:2}. The heat maps show that the functional correlation (or covariance) structure among the stocks is dense whereas that among the functional idiosyncratic components (after removing the common factors) is very sparse, justifying the proposed (approximate) functional low-rank plus sparse structure. Furthermore, we also note that the correlations among the stocks are stronger in 2020 than in 2019, which may be due to a stronger co-movement after the declaration of the COVID-19 pandemic in early 2020.

\section{Conclusion}\label{sec6}
\renewcommand{\theequation}{6.\arabic{equation}}
\setcounter{equation}{0}

In this paper, we introduce a general dual functional factor model framework to tackle high-dimensional functional time series, extending recent proposals in the literature \citep{GQW21, TNH23a, TNH23b}. The main model combines a high-dimensional fully functional factor model for the functional observations and a low-dimensional one for the latent functional factors. Using a sieve approximation for the functional factors, we approximate the large matrix of covariance functions (for the observed functional time series) by the functional low-rank plus sparse structure, making it feasible to extend \cite{FLM13}'s POET method to the large-scale functional data setting. A functional version of principal component analysis is proposed to estimate the functional factor loadings and common factors, and a functional shrinkage technique is adopted to estimate the covariance structure of the functional idiosyncratic components. Under some mild assumptions, we establish the convergence properties of the developed estimators. An amended information criterion is proposed to consistently estimate the factor number whereas a modified cross-validation is used to select the shrinkage parameter. Simulation study is provided to demonstrate reliable finite-sample performance of the developed methodology and the empirical application confirms the rationality of adopting the functional low-rank plus sparse covariance (or correlation) structure for CIDR curves of S\&P 500 index.


\section*{Acknowledgements}

Leng’s research is partly supported by EPSRC (EP/X009505/1). Li's research is partly supported by the Leverhulme Research Fellowship (RF-2023-396), Australian Research Council Discovery Project (DP230102250) and Heilbronn Institute for Mathematical Research. Shang's research is partly supported by the Australian Research Council Discovery Project (DP230102250). Xia's research is partly supported by the National Natural Science Foundation of China (72033002).


\section*{Appendix A:\ Proofs of the main results}
\renewcommand{\theequation}{A.\arabic{equation}}
\setcounter{equation}{0}

We next prove the main asymptotic theorems. Throughout the proofs, we let $M$ denote a generic positive constant whose value may change from line to line.

\medskip

\noindent{\bf Proof of Proposition \ref{prop:4.1}}.\ \ (i) It follows from (\ref{eq2.7}) that the $(t,s)$-entry of $\Delta$ defined in (\ref{eq3.5}) can be written as
\begin{eqnarray}
\Delta_{ts}&=&\frac{1}{N}\sum_{i=1}^N \int_{u\in{\mathbb C}_i}\left[\Lambda_i(u)^{^\intercal}G_t+\varepsilon_{it}^\ast(u)\right] \left[\Lambda_i(u)^{^\intercal}G_s+\varepsilon_{is}^\ast(u)\right]du\nonumber\\
&=&G_t^{^\intercal}\left[\frac{1}{N}\sum_{i=1}^N \int_{u\in{\mathbb C}_i}\Lambda_i(u) \Lambda_i(u)^{^\intercal}\right] G_s+\frac{1}{N}\sum_{i=1}^N \int_{u\in{\mathbb C}_i}G_s^{^\intercal}\Lambda_i(u)\varepsilon_{it}^\ast(u)du+\nonumber\\
&&\frac{1}{N}\sum_{i=1}^N \int_{u\in{\mathbb C}_i} G_t^{^\intercal}\Lambda_i(u)\varepsilon_{is}^\ast(u)du+\frac{1}{N}\sum_{i=1}^N \int_{u\in{\mathbb C}_i} \varepsilon_{it}^\ast(u)\varepsilon_{is}^\ast(u)du.\nonumber
\end{eqnarray}
Then, by the definition of the PCA estimate $\widetilde G_t$, we have
\begin{equation}\label{eqA.1}
\widetilde G_t-R G_t=V^{-1}\frac{1}{NT}\sum_{s=1}^{T}\sum_{i=1}^N\int_{u\in{\mathbb C}_i}\left[\widetilde{G}_s G_s^{^\intercal}\Lambda_i(u)\varepsilon_{it}^\ast(u)+\widetilde{G}_s \varepsilon_{is}^\ast(u)\Lambda_i(u)^{^\intercal}G_t+ \widetilde{G}_s\varepsilon_{is}^\ast(u)\varepsilon_{it}^\ast(u)\right]du,
\end{equation}
where $V$ and $R$ are defined in Section \ref{sec4.2}.

By the triangle and Cauchy-Schwarz inequalities and noting that $\varepsilon_{it}^\ast=\chi_{it}^\eta+\varepsilon_{it}$,
\begin{eqnarray}
&&\frac{1}{N^2T^3}\sum_{t=1}^T\left\vert\sum_{s=1}^{T}\sum_{i=1}^N\int_{u\in{\mathbb C}_i}\widetilde{G}_s G_s^{^\intercal}\Lambda_i(u)\varepsilon_{it}^\ast(u)du\right\vert_2^2\nonumber\\
&\leq&\frac{2}{N^2T^3}\sum_{t=1}^T\left\vert\sum_{s=1}^{T}\widetilde{G}_s G_s^{^\intercal} \sum_{i=1}^N\langle\Lambda_i, \varepsilon_{it}\rangle\right\vert_2^2+\frac{2}{N^2T^3}\sum_{t=1}^T\left\vert\sum_{s=1}^{T}\widetilde{G}_s G_s^{^\intercal}\sum_{i=1}^N\langle\Lambda_i, \chi_{it}^\eta\rangle\right\vert_2^2\nonumber\\
&\leq&2\left(\frac{1}{T}\sum_{s=1}^{T}\left\vert\widetilde{G}_s\right\vert_2^2\right)\left(\frac{1}{T}\sum_{s=1}^{T}\left\vert G_s\right\vert_2^2\right)\left(\frac{1}{N^2T} \sum_{t=1}^T\left\vert\sum_{i=1}^N\langle\Lambda_i, \varepsilon_{it}\rangle\right\vert_2^2\right)+\nonumber\\
&&2\left(\frac{1}{T}\sum_{s=1}^{T}\left\vert\widetilde{G}_s\right\vert_2^2\right)\left(\frac{1}{T}\sum_{s=1}^{T}\left\vert G_s\right\vert_2^2\right)\left(\frac{1}{N^2T} \sum_{t=1}^T\left\vert\sum_{i=1}^N\langle\Lambda_i, \chi_{it}^\eta\rangle\right\vert_2^2\right).\label{eqA.2}
\end{eqnarray}
By the identification condition (\ref{eq3.6}) and Assumption \ref{ass:1}(i), we have
\begin{equation}\label{eqA.3}
\frac{1}{T}\sum_{s=1}^{T}\left\vert\widetilde{G}_s\right\vert_2^2=O_P\left(q\right),\ \ \frac{1}{T}\sum_{s=1}^{T}\left\vert G_s\right\vert_2^2=O_P\left(q\right).
\end{equation}
By Assumptions \ref{ass:1}(ii) and \ref{ass:2}(ii),
\begin{eqnarray}
&&{\sf E}\left[\sum_{t=1}^T\left\vert\sum_{i=1}^N\langle\Lambda_i, \varepsilon_{it}\rangle\right\vert_2^2\right]\leq\sum_{t=1}^T{\sf E}\left[\left\vert\sum_{i=1}^N\langle\Lambda_i, \varepsilon_{it}\rangle\right\vert_2^2\right]\nonumber\\
&&\leq M\sum_{t=1}^T\sum_{i=1}^N\sum_{k=1}^q \Vert \Lambda_{ik}\Vert_2^2=O\left(TNq\right),\label{eqA.4}
\end{eqnarray}
where $\Lambda_{ik}(\cdot)$ is the $k$-th element of $\Lambda_i(\cdot)$. By Assumption \ref{ass:1}(ii)--(iii), we may show that ${\sf E}\left[\Vert \chi_{it}^\eta\Vert_2^2\right]=O\left(\xi_q^2\right)$ uniformly over $i$ and $t$, which, together with the Cauchy-Schwarz inequality, leads to
\begin{eqnarray}
&&{\sf E}\left[\sum_{t=1}^T\left\vert\sum_{i=1}^N\langle\Lambda_i, \chi_{it}^\eta\rangle\right\vert_2^2\right]\leq \left(\sum_{i=1}^N\sum_{k=1}^q \Vert \Lambda_{ik}\Vert_2^2\right)\left(\sum_{t=1}^T\sum_{i=1}^N {\sf E}\left[\Vert \chi_{it}^\eta\Vert_2^2\right]\right)\nonumber\\
&&\leq M(Nq)(NT\xi_q^2)=O\left(TN^2q\xi_q^2\right).\label{eqA.5}
\end{eqnarray}
With (\ref{eqA.2})--(\ref{eqA.5}), we readily have that
\begin{equation}\label{eqA.6}
\frac{1}{N^2T^3}\sum_{t=1}^T\left\vert\sum_{s=1}^{T}\sum_{i=1}^N\int_{u\in{\mathbb C}_i}\widetilde{G}_s G_s^{^\intercal}\Lambda_i(u)\varepsilon_{it}^\ast(u)du\right\vert_2^2=O_P\left(q^2\left(qN^{-1}+q\xi_q^2\right)\right).
\end{equation}

Similarly to (\ref{eqA.2}), we have
\begin{eqnarray}
&&\frac{1}{N^2T^3}\sum_{t=1}^T\left\vert\sum_{s=1}^{T}\sum_{i=1}^N\int_{u\in{\mathbb C}_i}\widetilde{G}_s \varepsilon_{is}^\ast(u)\Lambda_i(u)^{^\intercal}G_tdu\right\vert_2^2\nonumber\\
&\leq&\frac{2}{N^2T^3}\sum_{t=1}^T\left\vert\sum_{s=1}^{T}\widetilde{G}_s  \sum_{i=1}^N\langle\varepsilon_{is}, \Lambda_i^{^\intercal}\rangle G_t\right\vert_2+\frac{2}{N^2T^3}\sum_{t=1}^T\left\vert\sum_{s=1}^{T}\widetilde{G}_s \sum_{i=1}^N\langle \chi_{is}^\eta,\Lambda_i^{^\intercal}\rangle G_t\right\vert_2^2\nonumber\\
&\leq&2\left(\frac{1}{T}\sum_{s=1}^{T}\left\vert\widetilde{G}_s\right\vert_2^2\right)\left(\frac{1}{T}\sum_{t=1}^{T}\left\vert G_t\right\vert_2^2\right) \left(\frac{1}{N^2T}\sum_{s=1}^T\left\vert\sum_{i=1}^N\langle\Lambda_i, \varepsilon_{is}\rangle\right\vert_2^2\right)+\nonumber\\
&&2\left(\frac{1}{T}\sum_{s=1}^{T}\left\vert\widetilde{G}_s\right\vert_2^2\right)\left(\frac{1}{T}\sum_{t=1}^{T}\left\vert G_t\right\vert_2^2\right) \left(\frac{1}{N^2T}\sum_{s=1}^T\left\vert\sum_{i=1}^N\langle\Lambda_i, \chi_{is}^\eta\rangle\right\vert_2^2\right).\label{eqA.7}
\end{eqnarray}
It follows from (\ref{eqA.4}) and (\ref{eqA.5}) that
\begin{equation}\label{eqA.8}
\sum_{s=1}^T\left\vert\sum_{i=1}^N\langle\Lambda_i, \varepsilon_{is}\rangle\right\vert_2^2=O_P\left(TNq\right),\ \ \ \sum_{s=1}^T\left\vert\sum_{i=1}^N\langle\Lambda_i, \chi_{is}^\eta\rangle\right\vert_2^2=O_P\left(TN^2q\xi_q^2\right).
\end{equation}
With (\ref{eqA.3}), (\ref{eqA.7}) and (\ref{eqA.8}), we have
\begin{equation}\label{eqA.9}
\frac{1}{NT^2}\sum_{t=1}^T\left\vert\sum_{s=1}^{T}\sum_{i=1}^N\int_{u\in{\mathbb C}_i}\widetilde{G}_s \varepsilon_{is}^\ast(u)\Lambda_i(u)^{^\intercal}G_tdu\right\vert_2^2=O_P\left(q^2\left(qN^{-1}+q\xi_q^2\right)\right).
\end{equation}

By the Cauchy-Schwarz inequality,
\begin{eqnarray}
&&\frac{1}{N^2T^3}\sum_{t=1}^T\left\vert\sum_{s=1}^{T}\sum_{i=1}^N\int_{u\in{\mathbb C}_i}\widetilde{G}_s\varepsilon_{is}^\ast(u)\varepsilon_{it}^\ast(u)du\right\vert_2^2\nonumber\\
&\leq&\left(\frac{1}{T}\sum_{s=1}^{T}\left\vert\widetilde{G}_s\right\vert_2^2\right)\left(\frac{1}{N^2T^2} \sum_{t=1}^{T}\sum_{s=1}^T\left\vert\sum_{i=1}^N\langle\varepsilon_{is}^\ast, \varepsilon_{it}^\ast\rangle\right\vert^2\right)\nonumber\\
&\leq&4\left(\frac{1}{T}\sum_{s=1}^{T}\left\vert\widetilde{G}_s\right\vert_2^2\right)\left(\frac{1}{N^2T^2} \sum_{t=1}^{T}\sum_{s=1}^T\left\vert\sum_{i=1}^N\langle\varepsilon_{is}, \varepsilon_{it}\rangle\right\vert^2\right)\nonumber\\
&&4\left(\frac{1}{T}\sum_{s=1}^{T}\left\vert\widetilde{G}_s\right\vert_2^2\right)\left(\frac{1}{N^2T^2} \sum_{t=1}^{T}\sum_{s=1}^T\left\vert\sum_{i=1}^N\langle\chi_{is}^\eta, \varepsilon_{it}\rangle\right\vert^2\right)+\nonumber\\
&&4\left(\frac{1}{T}\sum_{s=1}^{T}\left\vert\widetilde{G}_s\right\vert_2^2\right)\left(\frac{1}{N^2T^2} \sum_{t=1}^{T}\sum_{s=1}^T\left\vert\sum_{i=1}^N\langle\varepsilon_{is}, \chi_{it}^\eta\rangle\right\vert^2\right)+\nonumber\\
&&4\left(\frac{1}{T}\sum_{s=1}^{T}\left\vert\widetilde{G}_s\right\vert_2^2\right)\left(\frac{1}{N^2T^2} \sum_{t=1}^{T}\sum_{s=1}^T\left\vert\sum_{i=1}^N\langle\chi_{is}^\eta, \chi_{it}^\eta\rangle\right\vert^2\right).\label{eqA.10}
\end{eqnarray}
By Assumption \ref{ass:2}(i), we readily have that
\begin{eqnarray}
\sum_{t=1}^{T}\sum_{s=1}^T\left\vert\sum_{i=1}^N\langle\varepsilon_{is}, \varepsilon_{it}\rangle\right\vert^2&\leq&2\left(\sum_{t=1}^{T}\sum_{s=1}^T\left\vert\sum_{i=1}^N\langle\varepsilon_{is}, \varepsilon_{it}\rangle-{\sf E}\left[\langle\varepsilon_{is}, \varepsilon_{it}\rangle\right] \right\vert^2\right)+\nonumber\\
&&2\left(\sum_{t=1}^{T}\sum_{s=1}^T\left\vert\sum_{i=1}^N{\sf E}\left[\langle\varepsilon_{is}, \varepsilon_{it}\rangle\right] \right\vert^2\right)\nonumber\\
&=&O_P\left(T^{2}N+N^2T\right).\label{eqA.11}
\end{eqnarray}
Since $\{\chi_{it}^\eta\}$ is independent of $\{\varepsilon_{it}\}$, by Assumption \ref{ass:2}(ii),
\[
\sum_{t=1}^T\sum_{s=1}^T{\sf E}\left(\left\vert\sum_{i=1}^N\langle\chi_{is}^\eta, \varepsilon_{it}\rangle\right\vert^2\right)\leq M\sum_{t=1}^T\sum_{s=1}^T\sum_{i=1}^N{\sf E}\left[\Vert \chi_{is}^\eta\Vert_2^2\right]=O\left(T^2N\xi_q^2\right),
\]
indicating that
\begin{equation}\label{eqA.12}
\sum_{t=1}^{T}\sum_{s=1}^T\left\vert\sum_{i=1}^N\langle\chi_{is}^\eta, \varepsilon_{it}\rangle\right\vert^2+\sum_{t=1}^{T}\sum_{s=1}^T\left\vert\sum_{i=1}^N\langle\varepsilon_{is}, \chi_{it}^\eta\rangle\right\vert^2=O_P\left(T^{2}N\xi_q^2\right).
\end{equation}
As ${\sf E}\left[\Vert \chi_{it}^\eta\Vert_2^2\right]=O\left(\xi_q^2\right)$, we can prove that
\begin{eqnarray}
\sum_{t=1}^{T}\sum_{s=1}^T\left\vert\sum_{i=1}^N\langle\chi_{is}^\eta, \chi_{it}^\eta\rangle\right\vert^2&\leq&\left(\sum_{s=1}^T\sum_{i=1}^N\Vert \chi_{is}^\eta\Vert_2^2\right) \left(\sum_{t=1}^{T}\sum_{i=1}^N\Vert\chi_{it}^\eta\Vert_2^2\right)\nonumber\\
&=&O_P\left(T^{2}N^2\xi_q^4\right).\label{eqA.13}
\end{eqnarray}
With (\ref{eqA.3}) and (\ref{eqA.10})--(\ref{eqA.13}),
\begin{equation}\label{eqA.14}
\frac{1}{NT^2}\sum_{t=1}^T\left\vert\sum_{s=1}^{T}\sum_{i=1}^N\int_{u\in{\mathbb C}_i}\widetilde{G}_s\varepsilon_{is}^\ast(u)\varepsilon_{it}^\ast(u)du\right\vert_2^2=O_P\left(q\left(N^{-1}+T^{-1}+\xi_q^4\right)\right).
\end{equation}

It follows from Lemma \ref{le:B.1} that the $q\times q$ matrix $V$ is positive definite with the minimum eigenvalue larger than a positive number {\em w.p.a.1}. Hence, the inverse of $V$ is well defined {\em w.p.a.1}. Then, by virtue of (\ref{eqA.1}), (\ref{eqA.6}), (\ref{eqA.9}) and (\ref{eqA.14}), we complete the proof of (\ref{eq4.1}).

\medskip

(ii) From the definition of $\widetilde\Lambda_i$, the factor model formulation (\ref{eq2.7}), and the normalisation restriction $\frac{1}{T}\sum_{t=1}^T\widetilde G_t\widetilde G_t^{^\intercal}=I_{q}$ in (\ref{eq3.6}), we have
\begin{eqnarray}
\widetilde \Lambda_i&=&\frac{1}{T}\sum_{t=1}^TX_{it}\widetilde G_t=\frac{1}{T}\sum_{t=1}^T\left(\chi_{it}+\chi_{it}^\eta+\varepsilon_{it}\right)\widetilde G_t\notag\\
&=&\frac{1}{T}\sum_{t=1}^T\widetilde G_t G_t^{^{\intercal}}\Lambda_i+\frac{1}{T}\sum_{t=1}^T\chi_{it}^\eta\widetilde G_t+\frac{1}{T}\sum_{t=1}^T\varepsilon_{it}\widetilde G_t\notag\\
&=&(R^{-1})^{^\intercal}\Lambda_i+\frac{1}{T}\sum_{t=1}^T\widetilde G_t\left(G_t-R^{-1}\widetilde G_t\right)^{^\intercal}\Lambda_i+\frac{1}{T}\sum_{t=1}^T\chi_{it}^\eta\widetilde G_t+\notag\\
&&\frac{R}{T}\sum_{t=1}^T\varepsilon_{it} G_t+\frac{1}{T}\sum_{t=1}^T\varepsilon_{it}\left(\widetilde G_t-RG_t\right).\label{eqA.15}
\end{eqnarray}

Noting that $\max\limits_{1\leq i\leq N}\max\limits_{1\leq k\leq k_\ast}\Vert {\mathbf B}_{ik}\Vert_O\leq m_B$ in Assumption \ref{ass:1}(ii), by the definition of $\Lambda_i$, we have
\begin{equation}\label{eqA.16}
\max_{1\leq i\leq N}||| \Lambda_i |||_2=O\left(q^{1/2}\right).
\end{equation}
With (\ref{eq4.1}), (\ref{eqA.3}) and (\ref{eqA.16}), we can prove that
\begin{eqnarray}
&&\max_{1\leq i\leq N}\Big{|}\Big{|}\Big{|} \frac{1}{T}\sum_{t=1}^T\widetilde G_t\left(G_t-R^{-1}\widetilde G_t\right)^{^\intercal}\Lambda_i \Big{|}\Big{|}\Big{|}_2\notag\\
&\leq&\left(\frac{1}{T}\sum_{t=1}^{T}\left\vert\widetilde{G}_t\right\vert_2^2\right)^{1/2} \left(\frac{1}{T}\sum_{t=1}^T\left\vert \widetilde G_t-R G_t\right\vert_2^2\right)^{1/2}\max_{1\leq i\leq N}||| \Lambda_i |||_2\notag\\
&=&O_P\left(q^{3/2}\left(T^{-1/2}+qN^{-1/2}+q\xi_q\right)\right).\label{eqA.17}
\end{eqnarray}
By Lemmas \ref{le:B.3} and \ref{le:B.4} in Appendix B, we have
\begin{equation}\label{eqA.18}
\max_{1\leq i\leq N}\Big{|}\Big{|}\Big{|} \frac{1}{T}\sum_{t=1}^T\chi_{it}^\eta\widetilde G_t\Big{|}\Big{|}\Big{|}_2\leq \left(\max_{1\leq i\leq N}\frac{1}{T}\sum_{t=1}^T\left\Vert\chi_{it}^\eta\right\Vert^2\right)^{1/2} \left(\frac{1}{T}\sum_{t=1}^{T}\left\vert\widetilde{G}_t\right\vert_2^2\right)^{1/2}=O_P\left(q^{1/2}\xi_q\right)
\end{equation}
and
\begin{equation}\label{eqA.19}
\max_{1\leq i\leq N}\Big{|}\Big{|}\Big{|}\frac{1}{T}\sum_{t=1}^T\varepsilon_{it} G_t\Big{|}\Big{|}\Big{|}_2=O_P\left(q^{1/2}T^{-1/2}\left[\log (N\vee T)\right]^{1/2+1/(2\gamma)}\right).
\end{equation}
By (\ref{eq4.1}) and Lemma \ref{le:B.5}, we may show that
\begin{eqnarray}
\max_{1\leq i\leq N}\Big{|}\Big{|}\Big{|}\frac{1}{T}\sum_{t=1}^T\varepsilon_{it}\left(\widetilde G_t-RG_t\right)\Big{|}\Big{|}\Big{|}_2&\leq& \left(\max_{1\leq i\leq N}\frac{1}{T}\sum_{t=1}^T\left\Vert\varepsilon_{it}\right\Vert^2\right)^{1/2}\left(\frac{1}{T}\sum_{t=1}^T\left\vert \widetilde G_t-R G_t\right\vert_2^2\right)^{1/2}\notag\\
&=&O_P\left(q^{1/2}\left(T^{-1/2}+qN^{-1/2}+q\xi_q\right)\right).\label{eqA.20}
\end{eqnarray}
By virtue of (\ref{eqA.15}) and (\ref{eqA.17})--(\ref{eqA.20}), we complete the proof of (\ref{eq4.2}).\hfill$\blacksquare$

\medskip

\noindent{\bf Proof of Theorem \ref{thm:4.2}}.\ \ Define
\[
\overline{C}_\varepsilon=\left(\overline{C}_{\varepsilon,ij}\right)_{N\times N}\ \ {\rm with}\ \
\overline{C}_{\varepsilon,ij}=\left(\overline C_{\varepsilon,ij}(u,v):\ u\in{\mathbb C}_i,\ v\in{\mathbb C}_j\right)=\frac{1}{T}\sum_{t=1}^T \varepsilon_{it}\varepsilon_{jt},\]
and recall that
\[
\widehat{C}_{\varepsilon,ij}=\left(\widehat C_{\varepsilon,ij}(u,v):\ u\in{\mathbb C}_i,\ v\in{\mathbb C}_j\right)=\frac{1}{T}\sum_{t=1}^T \widetilde\varepsilon_{it}\widetilde\varepsilon_{jt}.
\]
We first prove that
\begin{equation}\label{eqA.21}
\max_{1\leq i\leq N}\max_{1\leq j\leq N}\left\Vert \widehat{C}_{\varepsilon,ij}-C_{\varepsilon,ij}\right\Vert_{\rm S}=O_P\left(\delta_{N,T,q}\right).
\end{equation}
By Lemma \ref{le:B.5} in Appendix B, we have
\begin{equation}\label{eqA.22}
\max_{1\leq i\leq N}\max_{1\leq j\leq N}\left\Vert \overline{C}_{\varepsilon,ij}-C_{\varepsilon,ij}\right\Vert_S=O_P\left(T^{-1/2}\left[\log (N\vee T)\right]^{1/2+1/(2\gamma)}\right)=o_P\left(\delta_{N,T,q}\right).
\end{equation}
Hence, to prove (\ref{eqA.21}), we only need to show
\begin{equation}\label{eqA.23}
\max_{1\leq i\leq N}\max_{1\leq j\leq N}\left\Vert \widehat{C}_{\varepsilon,ij}-\overline C_{\varepsilon,ij}\right\Vert_{\rm S}=O_P\left(\delta_{N,T,q}\right).
\end{equation}
Note that
\begin{eqnarray}
\widehat{C}_{\varepsilon,ij}-\overline C_{\varepsilon,ij}&=&\frac{1}{T}\sum_{t=1}^T \widetilde\varepsilon_{it}\widetilde\varepsilon_{jt}-\frac{1}{T}\sum_{t=1}^T \varepsilon_{it}\varepsilon_{jt}\notag\\
&=&\frac{1}{T}\sum_{t=1}^T \left(\widetilde\varepsilon_{it}-\varepsilon_{it}\right)\varepsilon_{jt}-\frac{1}{T}\sum_{t=1}^T \varepsilon_{it}\left(\widetilde\varepsilon_{jt}-\varepsilon_{jt}\right)+\frac{1}{T}\sum_{t=1}^T \left(\widetilde\varepsilon_{it}-\varepsilon_{it}\right)\left(\widetilde\varepsilon_{jt}-\varepsilon_{jt}\right)\notag
\end{eqnarray}
and
\[
\widetilde\varepsilon_{it}-\varepsilon_{it}=X_{it}-\widetilde\Lambda_{i}^{^\intercal}\widetilde G_t-\varepsilon_{it}=\Lambda_{i}^{^\intercal} G_t-\widetilde\Lambda_{i}^{^\intercal}\widetilde G_t +\chi_{it}^\eta.
\]
By (\ref{eqA.3}) and Proposition \ref{prop:4.1}, we may show that
\begin{eqnarray}
&&\max_{1\leq i\leq N}\frac{1}{T}\sum_{t=1}^T\left\Vert \Lambda_{i}^{^\intercal} G_t-\widetilde\Lambda_{i}^{^\intercal}\widetilde G_t\right\Vert_2^2\notag\\
&\leq&\max_{1\leq i\leq N}\frac{2}{T}\sum_{t=1}^T\left\Vert \Lambda_{i}^{^\intercal}\left(G_t-R^{-1}\widetilde G_t\right)\right\Vert_2^2+\max_{1\leq i\leq N}\frac{2}{T}\sum_{t=1}^T\left\Vert \left[\left(R^{-1}\right)^{^\intercal}\Lambda_{i}-\widetilde\Lambda_{i}\right]^{^\intercal}\widetilde G_t\right\Vert_2^2\notag\\
&\leq&2\max_{1\leq i\leq N}||| \Lambda_i |||_2^2\left(\frac{1}{T}\sum_{t=1}^T\left\vert G_t-R^{-1}\widetilde G_t\right\vert_2^2\right)+2\max_{1\leq i\leq N}||| \left(R^{-1}\right)^{^\intercal}\Lambda_{i}-\widetilde\Lambda_{i} |||_2^2 \left(\frac{1}{T}\sum_{t=1}^T\left\vert \widetilde G_t\right\vert_2^2\right)\notag\\
&=&O_P\left(q^2\left(T^{-1}+q^2N^{-1}+q^2\xi_q^2\right)\right)+O_P\left(\delta_{N,T,q}^2\right)\notag\\
&=&O_P\left(\delta_{N,T,q}^2\right).\label{eqA.24}
\end{eqnarray}
Combining (\ref{eqA.24}) and Lemma \ref{le:B.3}, we readily have that
\begin{equation}\label{eqA.25}
\max_{1\leq i\leq N}\frac{1}{T}\sum_{t=1}^T \left\Vert\widetilde\varepsilon_{it}-\varepsilon_{it}\right\Vert_2^2=O_P\left(\delta_{N,T,q}^2+\xi_q^2\right)=O_P\left(\delta_{N,T,q}^2\right).
\end{equation}
By (\ref{eqA.25}), Lemma \ref{le:B.5} and the Cauchy-Schwarz inequality, we prove (\ref{eqA.23}), which, together with (\ref{eqA.22}), leads to (\ref{eqA.21}). Using (\ref{eqA.21}) and the property (iii) of the shrinkage function $s_\rho(\cdot)$ and noting that $\rho=m_\rho\delta_{N,T,q}$, we may show that
\begin{eqnarray}
\max_{1\leq i\leq N}\max_{1\leq j\leq N}\left\Vert \widetilde{C}_{\varepsilon,ij}-C_{\varepsilon,ij}\right\Vert_{\rm S}&\leq& \max_{1\leq i\leq N}\max_{1\leq j\leq N}\left\Vert \widetilde{C}_{\varepsilon,ij}-\widehat C_{\varepsilon,ij}\right\Vert_{\rm S}+\max_{1\leq i\leq N}\max_{1\leq j\leq N}\left\Vert \widehat{C}_{\varepsilon,ij}-C_{\varepsilon,ij}\right\Vert_{\rm S}\notag\\
&=&O_P(\rho)+O_P\left(\delta_{N,T,q}\right)=O_P\left(\delta_{N,T,q}\right),\notag
\end{eqnarray}
completing the proof of (\ref{eq4.3}).

We next turn to the proof of (\ref{eq4.4}). Note that
\begin{eqnarray}
\max_{1\leq i\leq N}\sum_{j=1}^N\left\Vert \widetilde{C}_{\varepsilon,ij}-C_{\varepsilon,ij}\right\Vert_{\rm S} &=&\max_{1\leq i\leq N}\sum_{j=1}^{N}\left\Vert s_{\rho}\left(\widehat{C}_{\varepsilon,ij}\right)I\left(\left\Vert \widehat{C}_{\varepsilon,ij}\right\Vert_S>\rho\right)-C_{\varepsilon,ij}\right\Vert_{\rm S}  \notag \\
&\le & \max_{1\leq i\leq N}\sum_{j=1}^{N}\left\Vert s_{\rho}\left(\widehat{C}_{\varepsilon,ij}\right)-\widehat{C}_{\varepsilon,ij}\right\Vert_{\rm S}I\left(\left\Vert \widehat{C}_{\varepsilon,ij}\right\Vert_{\rm S}>\rho\right)+  \notag
\\
& &\max_{1\leq i\leq N}\sum_{j=1}^{N}\left\Vert \widehat{C}_{\varepsilon,ij}-C_{\varepsilon,ij}\right\Vert_{\rm S}I\left(\left\Vert \widehat{C}_{\varepsilon,ij}\right\Vert_{\rm S}>\rho\right)+
\notag \\
&&\max_{1\leq i\leq N}\sum_{j=1}^{N}\left\Vert C_{\varepsilon,ij}\right\Vert_S
I\left(\left\Vert \widehat{C}_{\varepsilon,ij}\right\Vert_S\leq\rho\right) \notag
\\
& =:& \Pi_1+\Pi_2+\Pi_3.  \label{eqA.26}
\end{eqnarray}

Define the event ${\cal E}(m)=\left\{\max\limits_{1\leq i,j\leq N}\left\Vert \widehat{C}_{\varepsilon,ij}-C_{\varepsilon,ij}\right\Vert_{\rm S}\le m\cdot\delta_{N,T,q}\right\}$, where $m$ is a positive constant. For any small $\epsilon>0$, by (\ref{eqA.22}), there exists $m_\epsilon>0$ such that
\begin{equation}  \label{eqA.27}
\mathsf{P}\left({\cal E}(m_\epsilon)\right)\geq 1-\epsilon.
\end{equation}
By the sparsity condition (\ref{eq3.3}) on $C_\varepsilon$, setting $\rho=m_\rho\delta_{N,T,q}$ with $m_\rho\geq2m_\epsilon$ and using the property (iii) of the shrinkage function $s_\rho(\cdot)$, we have
\begin{eqnarray}
\Pi_1+\Pi_2 &\le&(m_\rho+m_\epsilon)\delta_{N,T,q}\left[\max_{1\leq i\leq N}\sum_{j=1}^{N}I\left(\left\Vert \widehat{C}_{\varepsilon,ij}\right\Vert_{\rm S}>\rho\right)\right]  \notag \\
&=& (m_\rho+m_\epsilon)\delta_{N,T,q}\left[\max_{1\leq i\leq N}\sum_{j=1}^{N}I\left(\left\Vert \widehat{C}_{\varepsilon,ij}\right\Vert_{\rm S}>m_\rho\delta_{N,T,q}\right)\right] \notag \\
&\le& (m_\rho+m_\epsilon)\delta_{N,T,q}\left[\max_{1\leq i\leq N}\sum_{j=1}^{N}I\left(\left\Vert C_{\varepsilon,ij}\right\Vert_{\rm S}>m_\epsilon \delta_{N,T,q}\right)\right] \notag \\
&=& O_P\left(\delta_{N,T,q}\right)\left[ \max_{1\leq i\leq N}\sum_{j=1}^{N}\frac{\left\Vert C_{\varepsilon,ij}\right\Vert_{\rm S}^\iota}{\left(m_\epsilon \delta_{N,T,q}\right)^\iota}\right]  \notag \\
&=&O_P\left(\varpi_N\delta_{N,T,q}^{1-\iota}\right)
 \label{eqA.28}
\end{eqnarray}
conditional on ${\cal E}(m_\epsilon)$.

Note that the events $\left\{\left\Vert \widehat C_{\varepsilon,ij}\right\Vert_{\rm S}\le \rho\right\}$ and ${\cal E}(m_\epsilon)$ jointly imply that $\left\{\left\Vert C_{\varepsilon,ij}\right\Vert_{\rm S}\le \left(m_\rho+m_\epsilon\right)\delta_{N,T,q}\right\}$ by the triangle inequality. Hence, conditional on ${\cal E}(m_\epsilon)$,
\begin{eqnarray}
\Pi_3 &\le&\max_{1\le i\le N}\sum_{j=1}^{N}\left\Vert C_{\varepsilon,ij}\right\Vert_{\rm S}
I\left(\left\Vert C_{\varepsilon,ij}\right\Vert_{\rm S}\le \left(m_\rho+m_\epsilon\right)\delta_{N,T,q}\right) \notag \\
&\le& \left(m_\rho+m_\epsilon\right)^{1-\iota}\delta_{N,T,q}^{1-\iota}\max_{1\le i\le N}\sum_{j=1}^{N}\left\Vert C_{\varepsilon,ij}\right\Vert_{\rm S}^\iota
\notag \\
&=&O_P\left(\varpi_N\delta_{N,T,q}^{1-\iota}\right).  \label{eqA.29}
\end{eqnarray}
By virtue of (\ref{eqA.28}) and (\ref{eqA.29}), letting $\epsilon\rightarrow0$ in (\ref{eqA.27}), we complete the proof of (\ref{eq4.4}).\hfill$\blacksquare$

\medskip

\noindent{\bf Proof of Theorem \ref{thm:4.3}}.\ As $\Sigma_G=I_q$, we have $C_{X,ij}^\ast=\Lambda_i^{^\intercal}\Lambda_j+C_{\varepsilon,ij}$, and thus
\[
\left\Vert \widetilde{C}_{X,ij}-C_{X,ij}^\ast\right\Vert_{\rm S}\leq  \left\Vert \widetilde{C}_{\varepsilon,ij}-C_{\varepsilon,ij}\right\Vert_{\rm S}+\left\Vert \widetilde\Lambda_i^{^\intercal}\widetilde\Lambda_j-\Lambda_i^{^\intercal}\Lambda_j\right\Vert_{\rm S}.
\]
By Proposition \ref{prop:4.1}(ii) and (\ref{eqA.16}), we may show that
\begin{equation}\label{eqA.30}
\max_{1\leq i\leq N}\max_{1\leq j\leq N}\left\Vert \widetilde\Lambda_i^{^\intercal}\widetilde\Lambda_j-\Lambda_i^{^\intercal}\Lambda_j\right\Vert_{\rm S}=O_P\left(q^{1/2}\delta_{N,T,q}\right).
\end{equation}
which, together with (\ref{eq4.3}) in Theorem \ref{thm:4.2}, leads to (\ref{eq4.5}).

Let $C_{\varepsilon,ij}^\ast$ be the covariance function between the functional idiosyncratic components $\varepsilon_{it}^\ast$ and $\varepsilon_{jt}^\ast$ defined in (\ref{eq2.7}). Following the proof of Lemma \ref{le:B.3} in Appendix B, we may show that
\[
\max_{1\leq i\leq N}\max_{1\leq j\leq N}\left\Vert C_{\varepsilon,ij}^\ast-C_{\varepsilon,ij}\right\Vert_{\rm S}=O_P\left(\xi_q\right)=o_P\left(\delta_{N,T,q}\right),
\]
which, together with (\ref{eq4.3}) and (\ref{eqA.30}), leads to
\begin{eqnarray}
\max_{1\leq i\leq N}\max_{1\leq j\leq N}\left\Vert \widetilde{C}_{X,ij}-C_{X,ij}\right\Vert_{\rm S}&\leq& \max_{1\leq i\leq N}\max_{1\leq j\leq N}\left[\left\Vert \widetilde{C}_{\varepsilon,ij}-C_{\varepsilon,ij}^\ast\right\Vert_{\rm S}+\left\Vert \widetilde\Lambda_i^{^\intercal}\widetilde\Lambda_j-\Lambda_i^{^\intercal}\Lambda_j\right\Vert_{\rm S}\right]\notag\\
&\leq&\max_{1\leq i\leq N}\max_{1\leq j\leq N}\left[ \left\Vert \widetilde{C}_{\varepsilon,ij}-C_{\varepsilon,ij}\right\Vert_{\rm S}+\right.\notag\\
&&\left.\left\Vert C_{\varepsilon,ij}^\ast-C_{\varepsilon,ij}\right\Vert_{\rm S}+\left\Vert \widetilde\Lambda_i^{^\intercal}\widetilde\Lambda_j-\Lambda_i^{^\intercal}\Lambda_j\right\Vert_{\rm S}\right]\notag\\
&=&O_P\left(q^{1/2}\delta_{N,T,q}\right),\notag
\end{eqnarray}
completing the proof of (\ref{eq4.6}).\hfill$\blacksquare$

\medskip

\noindent{\bf Proof of Proposition \ref{prop:5.1}}.\ By Lemma \ref{le:B.1} and the condition $\phi_{N,T}=o(1)$ in (\ref{eq5.2}), $\nu_k(\Delta/T)$ is the leading term and decreasing over $1\leq k\leq q$, whereas the penalty term dominates $\nu_k(\Delta/T)$ and is increasing over $q+1\leq k\leq q_{\max}$. Hence the objective function in (\ref{eq5.1}) achieves its minimum at $k=q+1$ and $\widetilde{q}=q$ {\em w.p.a.1}. \hfill$\blacksquare$


\section*{Appendix B:\ Proofs of technical lemmas}
\renewcommand{\theequation}{B.\arabic{equation}}
\setcounter{equation}{0}

In this appendix, we provide the proofs of the technical lemmas which are involved in proofs of the main theorems in Appendix A.

\renewcommand{\thelemma}{B.\arabic{lemma}}
\setcounter{lemma}{0}

\begin{lemma}\label{le:B.1}

Suppose that Assumptions \ref{ass:1}, \ref{ass:2} and \ref{ass:3}(i) are satisfied. The diagonal matrix $V$ with the diagonal elements being the first $q$ eigenvalues of $\frac{1}{T}\Delta$ (arranged in the decreasing order) is positive definite w.p.a.1.

\end{lemma}

\noindent{\bf Proof of Lemma \ref{le:B.1}}. Letting $\varepsilon_{i\bullet}^\ast=(\varepsilon_{i1}^\ast,\cdots,\varepsilon_{iT}^\ast)^{^\intercal}$, we note that
\begin{eqnarray}
\frac{1}{T}\Delta&=&\frac{1}{NT}G\left[\sum_{i=1}^N \int_{u\in{\mathbb C}_i}\Lambda_i(u)\Lambda_i(u)^{^\intercal}du\right]G^{^\intercal}+\frac{1}{NT}G\left[\sum_{i=1}^N \int_{u\in{\mathbb C}_i}\Lambda_i(u)\varepsilon_{i\bullet}^\ast(u)^{^\intercal}du\right]\nonumber\\
&&\frac{1}{NT}\left[\sum_{i=1}^N \int_{u\in{\mathbb C}_i}\varepsilon_{i\bullet}^\ast(u)\Lambda_i(u)^{^\intercal}du\right]G^{^\intercal}+\frac{1}{NT}\left[\sum_{i=1}^N \int_{u\in{\mathbb C}_i}\varepsilon_{i\bullet}^\ast(u)\varepsilon_{i\bullet}^\ast(u)^{^\intercal}du\right]\nonumber\\
&=:&\Delta_1+\Delta_2+\Delta_3+\Delta_4.\label{eqB.1}
\end{eqnarray}
As in the proofs of (\ref{eqA.6}) and (\ref{eqA.9}), we may show that
\begin{eqnarray}
\Vert \Delta_2\Vert&=&\left\Vert\frac{1}{NT}G\left[\sum_{i=1}^N \int_{u\in{\mathbb C}_i}\Lambda_i(u)\varepsilon_{i\bullet}^\ast(u)^{^\intercal}du\right]\right\Vert\nonumber\\
&\leq&2\left(\frac{1}{T}\sum_{t=1}^{T}\left\vert G_t\right\vert_2^2\right)^{1/2}\left[ \left(\frac{1}{N^2T}\sum_{s=1}^T\left\vert\sum_{i=1}^N\langle\Lambda_i, \varepsilon_{is}\rangle\right\vert_2^2\right)^{1/2}+\left(\frac{1}{N^2T}\sum_{s=1}^T\left\vert\sum_{i=1}^N\langle\Lambda_i, \chi_{is}^\eta\rangle\right\vert_2^2\right)^{1/2}\right]\nonumber\\
&=&O_P\left(q\left(N^{-1/2}+\xi_q\right)\right)=o_P(1),\label{eqB.2}
\end{eqnarray}
and
\begin{equation}\label{eqB.3}
\Vert \Delta_3\Vert=\left\Vert \frac{1}{NT}\left[\sum_{i=1}^N \int_{u\in{\mathbb C}_i}\varepsilon_{i\bullet}^\ast(u)\Lambda_i(u)^{^\intercal}du\right]G^{^\intercal}\right\Vert=O_P\left(q\left(N^{-1/2}+\xi_q\right)\right)=o_P(1).
\end{equation}
As in the proof of (\ref{eqA.14}), we have
\begin{equation}\label{eqB.4}
\Vert \Delta_4\Vert=\left\Vert \frac{1}{NT}\left[\sum_{i=1}^N \int_{u\in{\mathbb C}_i}\varepsilon_{i\bullet}^\ast(u)\varepsilon_{i\bullet}^\ast(u)^{^\intercal}du\right]\right\Vert=O_P\left(N^{-1/2}+T^{-1/2}+\xi_q^2\right)=o_P(1).
\end{equation}

Let $\nu_k(\Delta/T)$ and $\nu_k(\Delta_1)$ denote the $k$-th largest eigenvalue of $\Delta/T$ and $\Delta_1$, respectively. By Weyl's inequality and (\ref{eqB.1})--(\ref{eqB.4}), we can prove that
\[
\left\vert \nu_k(\Delta/T)-\nu_k(\Delta_1)\right\vert\leq \Vert \Delta_2\Vert+\Vert \Delta_3\Vert+\Vert \Delta_4\Vert=o_P(1)
\]
for any $k=1,\cdots,q$. By Assumption \ref{ass:1}(i)--(ii), $\nu_q(\Delta_1)$ is strictly positive. Hence, we may claim that $\nu_q(\Delta/T)$ is larger than a positive number {\em w.p.a.1}, completing the proof of Lemma \ref{le:B.1}. \hfill$\blacksquare$

\medskip

\begin{lemma}\label{le:B.2}

Suppose that Assumptions \ref{ass:1}, \ref{ass:2} and \ref{ass:3}(i) are satisfied. Let $V_0$ be a diagonal matrix with the diagonal elements being the eigenvalues of $\Sigma_\Lambda^{1/2}\Sigma_G\Sigma_\Lambda^{1/2}$ (arranged in the decreasing order) and $W_0$ be a matrix of the corresponding eigenvectors, where $\Sigma_\Lambda$ and $\Sigma_G$ are defined in Assumption \ref{ass:1}. Then, we have the following convergence result for the rotation matrix $R$:
\begin{equation}\label{eqB.5}
\left\Vert R-V_0^{-1/2}W_0\Sigma_\Lambda^{1/2}\right\Vert=o_P(1).
\end{equation}

\end{lemma}

\noindent{\bf Proof of Lemma \ref{le:B.2}}. Write
\[
\Sigma_{\Lambda,N}=\frac{1}{N}\sum_{i=1}^N \int_{u\in{\mathbb C}_i}\Lambda_i(u)\Lambda_i(u)^{^\intercal}du,\ \ \Sigma_{G,T}=\frac{1}{T} G^{^\intercal}G,\ \ \widetilde\Sigma_{G,T}=\frac{1}{T} G^{^\intercal}\widetilde G
\]
and define
\[\widetilde{W}=WD_W^{-1},\ \ W= \Sigma_{\Lambda,N}^{1/2}\widetilde\Sigma_{G,T},\ \ \ D_W=\left({\sf diag}\left\{W^{^\intercal}W\right\}\right)^{1/2},\]
where ${\sf diag}\{A\}$ denotes diagonalisation of a square matrix. Let
\[
\Omega_{N,T}= \Sigma_{\Lambda,N}^{1/2}\Sigma_{G,T}\Sigma_{\Lambda,N}^{1/2},\ \ \
\Omega_\ast=\Sigma_{\Lambda,N}^{1/2}\frac{G^{^\intercal}}{T^{1/2}}\left(\frac{\Delta}{T}-\Delta_1\right)\frac{\widetilde
G}{T^{1/2}},
\]
where $\Delta$ is defined in (\ref{eq3.5}) and $\Delta_1$ is defined in (\ref{eqB.1}). By the definition of the functional PCA estimation in Section \ref{sec3}, we have
\begin{equation}\label{eqB.6}
\left(\Omega_{N,T}+\Omega_\ast W^{-1}\right)\widetilde{W}=\widetilde{W}V,
\end{equation}
indicating that $\widetilde{W}$ is a $q\times q$ matrix consisting of the eigenvectors of $\Omega_{N,T}+\Omega_\ast W^{-1}$.

Rewrite $R=V^{-1}\left(D_W\widetilde{W}^{^\intercal}\right)\Sigma_{\Lambda,N}^{1/2}$. By Lemma \ref{le:B.1} and Assumption \ref{ass:1}(i)--(ii), in order to complete the proof of the lemma, we only need to show
\begin{equation}\label{eqB.7}
\left\Vert V-V_0\right\Vert=o_P(1),\ \ \left\Vert D_W^2-V_0\right\Vert=o_P(1),\ \ \
\left\Vert \widetilde W-W_0\right\Vert=o_P(1).
\end{equation}

By Lemma \ref{le:B.1}, we readily have the first assertion in (\ref{eqB.7}). By the triangle inequality, we have
\[
\left\Vert D_W^2-V_0\right\Vert\leq \left\Vert D_W^2-V\right\Vert+\left\Vert V-V_0\right\Vert=\left\Vert D_W^2-V\right\Vert+o_P(1).
\]
By the definitions of $D_W$ and functional PCA as well as Lemma \ref{le:B.1},
\begin{eqnarray}
\left\Vert D_W^2-V\right\Vert&\leq&\left\Vert \frac{1}{T}\widetilde G^{^\intercal}\left(\frac{1}{T}\Delta-\frac{1}{T}G \Sigma_{\Lambda,N}G^{^\intercal}\right)\widetilde G\right\Vert\notag\\
&\leq&\left\Vert\frac{1}{T}\Delta-\frac{1}{T}G \Sigma_{\Lambda,N}G^{^\intercal}\right\Vert\notag\\
&=&\left\Vert\frac{1}{T}\Delta-\Delta_1\right\Vert=o_P(1),\label{eqB.8}
\end{eqnarray}
completing the proof of the second assertion in (\ref{eqB.7}). We next turn to the proof of the third assertion in (\ref{eqB.7}). Let $W_{N,T}$ be a $q\times q$ matrix consisting of the eigenvectors of $\Omega_{N,T}$. By the triangle inequality again and Assumption \ref{ass:1}(ii)--(iii), we have
\begin{equation}\label{eqB.9}
\left\Vert \widetilde W-W_0\right\Vert\leq \left\Vert \widetilde W-W_{N,T}\right\Vert+\left\Vert  W_{N,T}-W_0\right\Vert=\left\Vert \widetilde W-W_{N,T}\right\Vert+o_P(1)
\end{equation}
By (\ref{eqB.6}), the definition of $W_{N,T}$ and the $\sin \theta$ theorem in \cite{DK70}, we have
\begin{equation}\label{eqB.10}
\left\Vert \widetilde W-W_{N,T}\right\Vert\leq M\cdot \Vert \Omega_\ast\Vert\cdot \Vert W^{-1}\Vert.
\end{equation}
Using Assumption \ref{ass:1}(i)--(ii) and Proposition \ref{prop:4.1}(i), and noting that the rotation matrix $R$ is asymptotically non-singular, we may show that $\Vert W^{-1}\Vert=O_P(1)$. By Lemma \ref{le:B.1}, we readily have $\Vert \Omega_\ast\Vert=o_P(1)$. Hence, we have $
\left\Vert \widetilde W-W_{N,T}\right\Vert=o_P(1)$, which, together with (\ref{eqB.10}), leads to the third assertion in (\ref{eqB.7}). The proof of Lemma \ref{le:B.2} is completed.\hfill$\blacksquare$

\begin{lemma}\label{le:B.3}

Suppose that Assumption \ref{ass:1}(iii) is satisfied and the factor loading operator ${\mathbf B}_{ik}$ in (\ref{eq2.1}) satisfies that $\max\limits_{1\leq i\leq N}\max\limits_{1\leq k\leq k_\ast}\Vert {\mathbf B}_{ik}\Vert_O\leq m_B<\infty$. Then we have
\begin{equation}\label{eqB.11}
\max_{1\leq i\leq N}\frac{1}{T}\sum_{t=1}^T\left\Vert\chi_{it}^\eta\right\Vert_2^2=O_P\left(\xi_q^2\right).
\end{equation}

\end{lemma}

\noindent{\bf Proof of Lemma \ref{le:B.3}}.\ By the definition of $\chi_{it}^\eta$, we have
\begin{eqnarray}
\max_{1\leq i\leq N}\left\Vert\chi_{it}^\eta\right\Vert_2&\leq& \max\limits_{1\leq i\leq N}\sum_{k=1}^{k_\ast}\Vert {\mathbf B}_{ik}\Vert_{\rm O}\Vert\eta_{tk}\Vert_2\notag\\
&\leq& k_\ast\max\limits_{1\leq i\leq N}\max\limits_{1\leq k\leq k_\ast}\Vert {\mathbf B}_{ik}\Vert_{\rm O}\max_{1\leq k\leq k_\ast}\Vert\eta_{tk}\Vert_2\notag\\
&\leq& k_\ast m_B\max_{1\leq k\leq k_\ast}\Vert\eta_{tk}\Vert_2,\notag
\end{eqnarray}
which indicates that
\[
\max_{1\leq i\leq N}\frac{1}{T}\sum_{t=1}^T\left\Vert\chi_{it}^\eta\right\Vert_2^2\leq (k_\ast m_B)\max_{1\leq k\leq k_\ast}\frac{1}{T}\sum_{t=1}^T\Vert\eta_{tk}\Vert_2^2.
\]
Hence, to prove (\ref{eqB.11}), we only need to show that
\begin{equation}\label{eqB.12}
\max_{1\leq k\leq k_\ast}\frac{1}{T}\sum_{t=1}^T\Vert\eta_{tk}\Vert_2^2=O_P\left(\xi_q^2\right)
\end{equation}
By Assumption \ref{ass:1}(iii) and the Markov inequality, we have, for $\kappa\rightarrow\infty$
\begin{eqnarray}
{\sf P}\left(\max_{1\leq k\leq k_\ast}\frac{1}{T}\sum_{t=1}^T\Vert\eta_{tk}\Vert_2^2>\kappa\xi_q\right)
&\leq&\sum_{k=1}^{k_\ast}{\sf P}\left(\frac{1}{T}\sum_{t=1}^T\Vert\eta_{tk}\Vert_2^2>\kappa\xi_q^2\right)\notag\\
&=&\left(\kappa\xi_q^2\right)^{-1}\sum_{k=1}^{k_\ast}\frac{1}{T}\sum_{t=1}^T {\sf E}\left[\Vert\eta_{tk}\Vert_2^2\right]\rightarrow0,\notag
\end{eqnarray}
completing the proof of (\ref{eqB.12}).\hfill$\blacksquare$

\begin{lemma}\label{le:B.4}

Suppose that Assumption \ref{ass:3} is satisfied, and ${\sf E}[\varepsilon_{it} G_{tk}]=0$ for any $k=1,\cdots,q$ and $i=1,\cdots,N$. Then we have
\begin{equation}\label{eqB.13}
\max_{1\leq i\leq N}\max_{1\leq k\leq q}\left\Vert\frac{1}{T}\sum_{t=1}^T\varepsilon_{it} G_{tk}\right\Vert_2=O_P\left(T^{-1/2}\left[\log (N\vee T)\right]^{1/2+1/(2\gamma)}\right),
\end{equation}
where $\gamma>0$ is defined in Assumption \ref{ass:3}(ii).

\end{lemma}

\noindent{\bf Proof of Lemma \ref{le:B.4}}.\ \ We first prove that
\begin{equation}\label{eqB.14}
{\sf P}\left(\max_{1\leq t\leq T}\max_{1\leq k\leq q} |G_{tk}|>\sqrt{(2/\nu_0)\log T}\right)\rightarrow0
\end{equation}
and
\begin{equation}\label{eqB.15}
{\sf P}\left(\max_{1\leq i\leq N}\max_{1\leq t\leq T}\Vert \varepsilon_{it}\Vert_2>\sqrt{(3/\nu_0)\log (N\vee T)}\right)\rightarrow0.
\end{equation}
In fact, by Assumption \ref{ass:3}(iii) and the Bonferroni and Markov inequalities, we may show that
\begin{eqnarray}
{\sf P}\left(\max_{1\leq t\leq T}\max_{1\leq k\leq q} |G_{tk}|>\sqrt{(2/\nu_0)\log T}\right)&\leq&\sum_{t=1}^T\sum_{k=1}^q {\sf P}\left( |G_{tk}|>\sqrt{(2/\nu_0)\log T}\right)\notag\\
&=&\sum_{t=1}^T\sum_{k=1}^q {\sf P}\left(\exp\left\{\nu_0G_{tk}^2\right\}>T^2\right)\notag\\
&\leq&\sum_{t=1}^T\sum_{k=1}^qT^{-2}{\sf E}\left[\exp\left\{\nu_0G_{tk}^2\right\}\right]\notag\\
&=&O\left(q/T\right)=o(1)\notag
\end{eqnarray}
since $q=o(T)$ by Assumption \ref{ass:3}(i), and
\begin{eqnarray}
{\sf P}\left(\max_{1\leq i\leq N}\max_{1\leq t\leq T}\Vert \varepsilon_{it}\Vert_2>\sqrt{(3/\nu_0)\log (N\vee T)}\right)&\leq&\sum_{i=1}^N\sum_{t=1}^T {\sf P}\left( \Vert \varepsilon_{it}\Vert_2>\sqrt{(3/\nu_0)\log (N\vee T)}\right)\notag\\
&=&\sum_{i=1}^N\sum_{t=1}^T {\sf P}\left(\exp\left\{\nu_0 \Vert \varepsilon_{it}\Vert_2^2\right\}>(N\vee T)^3\right)\notag\\
&\leq&\sum_{i=1}^N\sum_{t=1}^T (N\vee T)^{-3}{\sf E}\left[\exp\left\{\nu_0 \Vert \varepsilon_{it}\Vert_2^2\right\}\right]\notag\\
&=&O\left((N\vee T)^{-1}\right)=o(1).\notag
\end{eqnarray}

By (\ref{eqB.14}) and (\ref{eqB.15}), without loss of generality, we assume that $|G_{tk}|\leq \sqrt{(2/\nu_0)\log T}$ and $\Vert \varepsilon_{it}\Vert_2\leq\sqrt{(3/\nu_0)\log (N\vee T)}$ in the rest of the proof. Hence, we have
\[
\max_{1\leq i\leq N}\max_{1\leq k\leq q}\max_{1\leq t\leq T}\Vert\varepsilon_{it} G_{tk}\Vert_2\leq c_{NT}:=(3/\nu_0)\log (N\vee T).
\]
It follows from Assumption \ref{ass:3}(iii) that there exists $0<\sigma_{\varepsilon G}^2<\infty$ such that
 \[
\max_{1\leq i\leq N}\max_{1\leq k\leq q}{\sf E}\left[\Vert\varepsilon_{it} G_{tk}\Vert_2^2\right]\leq \sigma_{\varepsilon G}^2.
\]
By Assumption \ref{ass:3}(ii), using the argument in the proof of Lemma S2.13 in \cite{TNH23b}, we may show that $\{\varepsilon_{it} G_{tk}: t=1,2,\cdots\}$ is a stationary sequence of ${\mathscr H}_i$-valued random elements with
\[
\tau(n)\leq\theta_{NT}\exp\left\{-(\theta_2n)^\gamma\right\},\ \ \theta_{NT}=\theta_1\sqrt{(10/\nu_0)\log(N\vee T)}.
\]
We next make use of the concentration inequality in Proposition \ref{prop:C.1} to prove (\ref{eqB.13}). Setting $v=2\log(N\vee T)$ and replacing $c_Y$ and $\sigma_Y$ by $c_{NT}$ and $\sigma_{\varepsilon G}$, respectively, we have
\begin{equation}\label{eqB.16}
v_{NT}=\frac{M_1(\sigma_{\varepsilon G},2\log(N\vee T))}{T_\dagger^{1/2}}+\frac{M_2(c_{NT},2\log(N\vee T))}{T_\dagger},
\end{equation}
where $M_1(\cdot,\cdot)$ and $M_2(\cdot,\cdot)$ are defined in Proposition \ref{prop:C.1} and
\[
T_\dagger=\left\lfloor(T/2)\theta_2\left(1\vee \log(c_{NT}^{-1}\theta_{NT}\theta_2T)\right)^{-1/\gamma}\right\rfloor\propto T(\log T)^{-1/\gamma}.
\]
Then, using Proposition \ref{prop:C.1}, we have
\begin{eqnarray}
&&{\sf P}\left(\max_{1\leq i\leq N}\max_{1\leq k\leq q}\left\Vert\frac{1}{T}\sum_{t=1}^T\varepsilon_{it} G_{tk}\right\Vert_2>v_{NT}\right)\notag\\
&\leq&\sum_{i=1}^N\sum_{k=1}^q {\sf P}\left(\left\Vert\frac{1}{T}\sum_{t=1}^T\varepsilon_{it} G_{tk}\right\Vert_2>v_{NT}\right)\notag\\
&\leq&2\sum_{i=1}^N\sum_{k=1}^q \exp\{-2\log(N\vee T)\}\notag\\
&=&O\left(Nq(N\vee T)^{-2}\right)=o(1)\label{eqB.17}
\end{eqnarray}
as $q=o(N\vee T)$ by Assumption \ref{ass:3}(i). Since $[\log(N\vee T)]^{3+1/\gamma}=o(T)$, the first term on the right side of (\ref{eqB.16}) is the leading term of $v_{NT}$ and furthermore,
\[
v_{NT}\propto T^{-1/2}\left[\log (N\vee T)\right]^{1/2+1/(2\gamma)},
\]
which, together with (\ref{eqB.17}), completes the proof of (\ref{eqB.13}).\hfill$\blacksquare$

\begin{lemma}\label{le:B.5}

Suppose that Assumption \ref{ass:3} is satisfied. Then we have
\begin{equation}\label{eqB.18}
\max_{1\leq i\leq N}\max_{1\leq j\leq N}\left\Vert\frac{1}{T}\sum_{t=1}^T\varepsilon_{it} \varepsilon_{jt}-C_{\varepsilon,ij}\right\Vert_{\rm S}=O_P\left(T^{-1/2}\left[\log (N\vee T)\right]^{1/2+1/(2\gamma)}\right),
\end{equation}
where $C_{\varepsilon,ij}$ is the covariance function between $\varepsilon_{it}$ and $\varepsilon_{jt}$.

\end{lemma}

\noindent{\bf Proof of Lemma \ref{le:B.5}}.\ The proof is similar to the proof of Lemma \ref{le:B.4}. By (\ref{eqB.15}), without loss of generality, we may assume that $\Vert \varepsilon_{it}\Vert_2\leq\sqrt{(3/\nu_0)\log (N\vee T)}$ uniformly over $i$ and $t$ throughout the proof, and thus
\begin{eqnarray}
\max_{1\leq i\leq N}\max_{1\leq j\leq N}\max_{1\leq t\leq T}\Vert\varepsilon_{it} \varepsilon_{jt}\Vert_{\rm S}&\leq& \max_{1\leq i\leq N}\max_{1\leq t\leq T}\Vert\varepsilon_{it}\Vert_2\max_{1\leq j\leq N}\max_{1\leq t\leq T}\Vert\varepsilon_{jt}\Vert_2\notag\\
&\leq& c_{NT}=(3/\nu_0)\log (N\vee T).\notag
\end{eqnarray}
By Assumption \ref{ass:3}(iii), there exists $0<\sigma_{\varepsilon}^2<\infty$ such that
 \[
\max_{1\leq i\leq N}\max_{1\leq j\leq N}{\sf E}\left[\Vert \varepsilon_{it} \varepsilon_{jt}-C_{\varepsilon,ij}\Vert_{\rm S}^2\right]\leq \sigma_{\varepsilon}^2.
\]
By Assumption \ref{ass:3}(ii), following the proof of Lemma S2.13 in \cite{TNH23b}, $\{\varepsilon_{it} \varepsilon_{jt}: t=1,2,\cdots\}$ is a stationary sequence of ${\mathscr H}_i\times {\mathscr H}_j$-valued random functions with
\[
\tau(n)=\theta_{NT}^\ddagger\exp\left\{-(\theta_2n)^\gamma\right\},\ \ \theta_{NT}^\ddagger=\theta_1\sqrt{(12/\nu_0)\log(N\vee T)}.
\]
As in (\ref{eqB.16}), we write
\begin{equation}\label{eqB.19}
v_{NT}^\ddagger=\frac{M_1(\sigma_{\varepsilon}, 3\log(N\vee T))}{T_\ddagger^{1/2}}+\frac{M_2(c_{NT},3\log(N\vee T))}{T_\ddagger},
\end{equation}
where we set $v=3\log(N\vee T)$ in Proposition \ref{prop:C.1}, and
\[
T_\ddagger=\left\lfloor(T/2)\theta_2\left(1\vee \log(c_{NT}^{-1}\theta_{NT}^\ddagger\theta_2T)\right)^{-1/\gamma}\right\rfloor\propto T(\log T)^{-1/\gamma}.
\]
By Proposition \ref{prop:C.1}, we may show that
\begin{eqnarray}
&&{\sf P}\left(\max_{1\leq i\leq N}\max_{1\leq j\leq N}\left\Vert\frac{1}{T}\sum_{t=1}^T\varepsilon_{it} \varepsilon_{jt}-C_{\varepsilon,ij}\right\Vert_{\rm S}>v_{NT}^\ddagger\right)\notag\\
&\leq&\sum_{i=1}^N\sum_{j=1}^N {\sf P}\left(\left\Vert\frac{1}{T}\sum_{t=1}^T\varepsilon_{it} \varepsilon_{jt}-C_{\varepsilon,ij}\right\Vert_{\rm S}>v_{NT}^\ddagger\right)\notag\\
&\leq&2\sum_{i=1}^N\sum_{j=1}^N \exp\{-3\log(N\vee T)\}\notag\\
&=&O\left(NT(N\vee T)^{-3}\right)=o(1).\label{eqB.20}
\end{eqnarray}
Noting that $v_{NT}^\ddagger\propto T^{-1/2}\left[\log (N\vee T)\right]^{1/2+1/(2\gamma)}$, by (\ref{eqB.20}), we complete the proof of (\ref{eqB.18}).\hfill$\blacksquare$

\bigskip


\section*{Appendix C:\ $\tau$-mixing dependence and concentration inequality}
\renewcommand{\theequation}{C.\arabic{equation}}
\setcounter{equation}{0}

In this appendix, we introduce the so-called $\tau$-mixing dependence for random elements defined in a separable Hilbert space ${\mathscr H}$ with norm $\Vert\cdot\Vert_{\mathscr H}$, and provide the relevant concentration inequality which plays a crucial role in the main theoretical proofs. The $\tau$-mixing dependence concept is introduced by \cite{DDLLLP07} and \cite{W10} for real-valued random processes. \cite{BZ19} extend the concept to Banach-valued random processes and derive some concentration inequalities. We next briefly review some definitions and results closely related to our model setting and assumptions.

Let $(\Omega, {\cal F}, P)$ be the probability space and ${\mathscr L}_p={\mathscr L}_p(\Omega, {\cal F}, P)$ be a $p$-integrable real function space with $\Vert\cdot\Vert_{{\mathscr L}_p}$ being the norm. Consider $\{Y_t\}$ as a sequence of random elements in ${\mathscr H}$. Define the following concept of $\tau$-mixing coefficient:
\[
\tau(k)=\sup\left\{\left\vert {\sf E}\left[Z\psi(Y_{t+k})\right]-{\sf E}[Z] {\sf E}\left[\psi(Y_{t+k})\right]\right\vert:\ t\geq1,\ Z\in {\mathscr L}_1(\Omega, {\cal F}_t^Y, P),\ {\sf E}\left[\left\Vert Z\right\Vert_{{\mathscr L}_1}\right]\leq1,\ \psi\in{\cal L}_{1}\right\},
\]
where ${\cal F}_t^Y$ is the $\sigma$-algebra generated by $Y_1,\cdots,Y_t$, and
\[
{\mathscr L}_1=\left\{\psi: {\cal B}\rightarrow{\cal R},\ \left\vert\psi(x)-\psi(y)\right\vert\leq \Vert x-y\Vert_{\mathscr H}\right\}
\]
with ${\cal B}\subset {\mathscr H}$ being a ball. The ${\mathscr H}$-valued process $\{Y_t\}$ is $\tau$-mixing dependent if $\tau(k)\rightarrow0$ as $k\rightarrow\infty$. The following concentration inequality follows from Proposition 3.6 and Corollary 3.7 in \cite{BZ19}.

\renewcommand{\theprop}{C.\arabic{prop}}
\setcounter{prop}{0}

\begin{prop}\label{prop:C.1}

{\em Suppose that $\{Y_t\}$ is a stationary sequence of ${\mathscr H}$-valued random elements with mean zero, $\Vert Y_1\Vert_{\mathscr H}\leq c_Y$, ${\sf E}\left[\Vert Y_1\Vert_{\mathscr H}^2\right]\leq \sigma_Y^2$, and
\[
\tau(k)\leq\theta_1\exp\left\{-(\theta_2k)^\gamma\right\},\ \ \theta_1,\theta_2,\gamma>0.
\]
Then, we have
\[
{\sf P}\left(
\left\Vert \frac{1}{T}\sum_{t=1}^TY_t\right\Vert_{\mathscr H}\geq \frac{M_1(\sigma_Y,v)}{T_\ast^{1/2}}+\frac{M_2(c_Y,v)}{T_\ast}\right)\leq 2\exp\{-v\},
\]
where $v>0$, $M_1(\sigma_Y,v)=\sigma_Y\left(4+6\sqrt{2v}\right)$, $M_2(c_Y,v)=c_Y\left\{4+\left[2+2\sqrt{2}(1+2/c_Y)\right]v\right\}$, and
\[
T_\ast=\left\lfloor(T/2)\theta_2\left(1\vee \log(c_Y^{-1}\theta_1\theta_2T)\right)^{-1/\gamma}\right\rfloor.
\]
}

\end{prop}

\bigskip
\bigskip


\end{document}